\documentclass[reprint,aps,amsmath,amssymb,twocolumn,prx,floats,footinbib,longbibliography,superscriptaddress]{revtex4-2} 
\usepackage{graphics,graphicx,epsfig, sidecap} 
\usepackage{color} 
\usepackage{epsf,epstopdf,wrapfig} 

\usepackage{textgreek} 
\usepackage{sidecap} 
\usepackage{bbm}
\usepackage{hyperref}

\newcommand{\beq}{\begin{equation}} 
\newcommand{\eeq}{\end{equation}} 
\newcommand{\beqn}{\begin{eqnarray}} 
\newcommand{\eeqn}{\end{eqnarray}}

\def\eqref#1{Eq.~(\ref{#1})}
\def\figref#1{Fig.~\ref{#1}}

\begin{document}

\title{Maximum entropy models for patterns of gene expression}

\author{Camilla Sarra}
\thanks{Corresponding author: \href{mailto:csarra@princeton.edu}{csarra@princeton.edu}}
\affiliation{Joseph Henry Laboratories of Physics}

\author{Leopoldo Sarra}
\affiliation{Flatiron Institute, 162 5th Avenue, New York, New York 10010 USA}

\author{Luca Di Carlo}
\affiliation{Joseph Henry Laboratories of Physics}
\affiliation{Lewis--Sigler Institute for Integrative Genomics}

\author{Trevor GrandPre}
\affiliation{Joseph Henry Laboratories of Physics}
\affiliation{Lewis--Sigler Institute for Integrative Genomics}
\affiliation{Princeton Center for Theoretical Science\\
    Princeton University, Princeton, NJ 08544 USA}

\author{Yaojun Zhang}
\affiliation{Joseph Henry Laboratories of Physics}
\affiliation{Lewis--Sigler Institute for Integrative Genomics}
\affiliation{Department of Physics and Astronomy and Department of Biophysics\\ John Hopkins University}

\author{Curtis G.~Callan, Jr.}
\affiliation{Joseph Henry Laboratories of Physics}
\affiliation{Laboratoire de Physique de l’\'Ecole Normale Sup\'erieure,
    F-75005 Paris, France}

\author{William Bialek}
\affiliation{Joseph Henry Laboratories of Physics}
\affiliation{Lewis--Sigler Institute for Integrative Genomics}

\begin{abstract}
    New experimental methods make it possible to measure the expression levels of many genes, simultaneously, in snapshots from thousands or even millions of individual cells.  Current approaches to analyze these experiments involve clustering or low-dimensional projections.
    Here we use the principle of maximum entropy to obtain a probabilistic description that captures the observed presence or absence of mRNAs from hundreds of genes in cells from the mammalian brain.
    We construct the Ising model compatible with experimental means and pairwise correlations, and validate it by showing that it gives good predictions for higher-order statistics.
    We notice that the probability distribution of cell states has many local maxima. By labeling cell states according to the associated maximum, we obtain a cell classification that agrees well with previous results that use traditional clustering techniques.
    Our results provide quantitative descriptions of gene expression statistics and interpretable criteria for defining cell classes, supporting the hypothesis that cell classes emerge from the collective interaction of gene expression levels.
\end{abstract}

\maketitle

\section{Introduction}
\label{introduction}

The many cells in a complex organism all contain essentially the same DNA; what distinguishes different cells is how the different genes are expressed, that is how many copies of each possible protein are made ~\cite{chaffey2003alberts}. Every step that leads from gene to messenger RNA (mRNA) to protein is regulated, and many of the same regulatory mechanisms are operative in unicellular organisms as they move through their life cycles.  It is reasonable to think of expression levels---the number of copies of each mRNA or the corresponding protein---as defining the state of the cell, although long term differences in expression also lead to changes in cell shape and organization.   One might further hope that these states are organized into classes, defining the different types of cell in the body, and that this classification of molecular states would agree with other classification schemes based on morphology and function ~\cite{vickaryous2006human,insel2013nih,kepecs2014interneuron,armananzas2015towards,arendt2016origin,zeng2017neuronal,cembrowski2018continuous,zeng2022cell}.

Current view of cell states and cell types is being revolutionized by experimental methods that make it possible to (almost) count every single mRNA in a cell, labelled by the gene from which it has been transcribed.  One approach is single-cell RNA sequencing (scRNAseq) ~\cite{tang2009mrna,treutlein+al_14,klein+al_15,macosko+al_15}: individual cells are manipulated using microfluidics, mRNA is purified and reverse transcribed into DNA, and these DNA molecules then are sequenced.  An alternative is multiplexed error--robust fluorescence in situ hybridization (MERFISH) ~\cite{lubeck+cai_12,chen+al_15}: cells are fixed and labeled by  fluorescently tagged DNA molecules that are complementary to the sequences of different mRNAs, and super--resolution microscopy is used to count the molecules.  The scRNAseq method interrogates all genes at once, while the optical methods such as MERFISH target a large number of particular genes by using multiple rounds of combinatorial labelling and imaging; the different methods also have different sources of error.

If the state of a cell is defined by the expression levels of hundreds or even thousands of genes, then ``state'' is a point in a very high--dimensional space. The dominant strategies for analyzing expression patterns across large numbers of genes have involved searching for low--dimensional projections in which the classes or types of cells become obvious ~\cite{ amir2013visne, vzurauskiene2016pcareduce, mcinnes2018umap, becht2019dimensionality}.  These projection methods have been supplemented by various clustering algorithms ~\cite{petegrosso2020machine, yu2022benchmarking}.

Here we take a different point of view, trying to construct a statistical mechanics of cell states.  Concretely, we are searching for a good approximation to the joint probability distribution of expression levels, analogous to the Boltzmann distribution over states in a system with many degrees of freedom.  In such a distribution, classes or types of cells should be visible as multiple distinct peaks in the distribution.  In contrast, the states traversed during the cell cycle would appear as a ridge in the distribution.

We could approach the distribution over cell states by writing simplified dynamical models, in the spirit of Kauffman's model for genetic networks ~\cite{kauffman_69a,kauffman_69b} or Hopfield's model for neural networks ~\cite{hopfield1982neural,hopfield1984}. Here we want to take advantage of new, larger scale experiments, so we use the maximum entropy method ~\cite{jaynes1982rationale}.  We recall that this method provides the least structured distribution that is consistent with some expectation values that can be measured reliably in the data.  Maximum entropy has been applied successfully in describing emergent phenomena in many different living systems: patterns of activity in networks of neurons ~\cite{schneidman2006weak,granot2013stimulus,tkavcik2014searching,meshulam2017collective}, evolution of sequences in protein families ~\cite{bialek+ranganathan_07,weigt+al_09,mora2010maximum,marks2011protein,morcos2011direct,bitbol2016inferring}, the propagation of order in flocks of birds ~\cite{bialek2012statistical,mora+al_16}, and more ~\cite{locasale2009maximum,banavar+al_10,demartino+al_18}.

As an example, we consider recent experiments using MERFISH to count 500 different species of mRNA in millions of cells from adult mouse brains ~\cite{yao2023high}; these data are part of the Allen Brain Cell Atlas ~\cite{Allen}.   We will see that much of the collective behavior is preserved in a binary representation that simply indicates whether each gene has zero or non--zero mRNA counts in a cell.  We build Ising models for these binary variables that match their pairwise correlations, and find that these models predict higher-order correlations with reasonable accuracy and no free parameters.  These models also predict that the probability distribution has many distinguishable peaks, corresponding to local minima of the effective energy function, and these can be placed in good correspondence with major classes of cells identified through other experiments.  We can go further, building separate Ising models for each class of cells, and find that class assignments of unlabeled cells based on these models reach the same level of precision as a classifier built with a neural network.

We conclude that relatively simple models, grounded in statistical physics, provide quantitative descriptions of the patterns of gene expression and interpretable guides to the definition of cell classes.  These results encourage us to take seriously the idea that cell classes or types  emerge from interacting genetic networks in the same way that other collective phenomena emerge in the thermodynamic limit.

\section{Experimental Data}

The problem of classifying cells is especially interesting in the brain, where the diversity of structures, functions, and expression patterns is dramatically larger than in other organs; the most recent estimates are that there may be several thousand distinguishable cell types ~\cite{conroy_23}.   In an effort to standardize the exploration of this complexity, the US National Institutes of Health launched the BRAIN Initiative Cell Consensus Network (BICCN), which builds atlases that integrate information about physiology, morphology, connectivity, and gene expression ~\cite{BICCN}. The Allen Brain Cell Atlas is one component of this project.

The Allen Brain Cell Atlas includes two steps of measurements ~\cite{yao2023high,Allen}.  In the first step, scRNA-seq was used to estimate the number of copies of $\sim 8000$ mRNA species in a total of $\sim 7\times 10^6$ cells coming from the brains of $\sim 300$ mice.  The cells then were grouped using clustering algorithms based on both the Euclidean distance and cosine similarity of the high--dimensional gene expression patterns.  This produced a reference taxonomy with a hierarchical structure with 338 major clusters, 34 classes and 7 divisions.

In a second step, $N_g = 500$ genes were chosen because their mRNA counts seem to have the greatest power in distinguishing clusters and classes.  These genes then were targeted in MERFISH experiments on $N_c \sim 4\times 10^6$ cells, from 59 coronal sections at $200\,\mu{\rm m}$ intervals in the entire brain. These cells were labeled using the previous taxonomy by mapping each cell to the closest cluster. Information about the locations and spatial relations of the cells, which are preserved in MERFISH, was in turn used to annotate the clusters.

\begin{figure}
    \includegraphics[width=\linewidth]{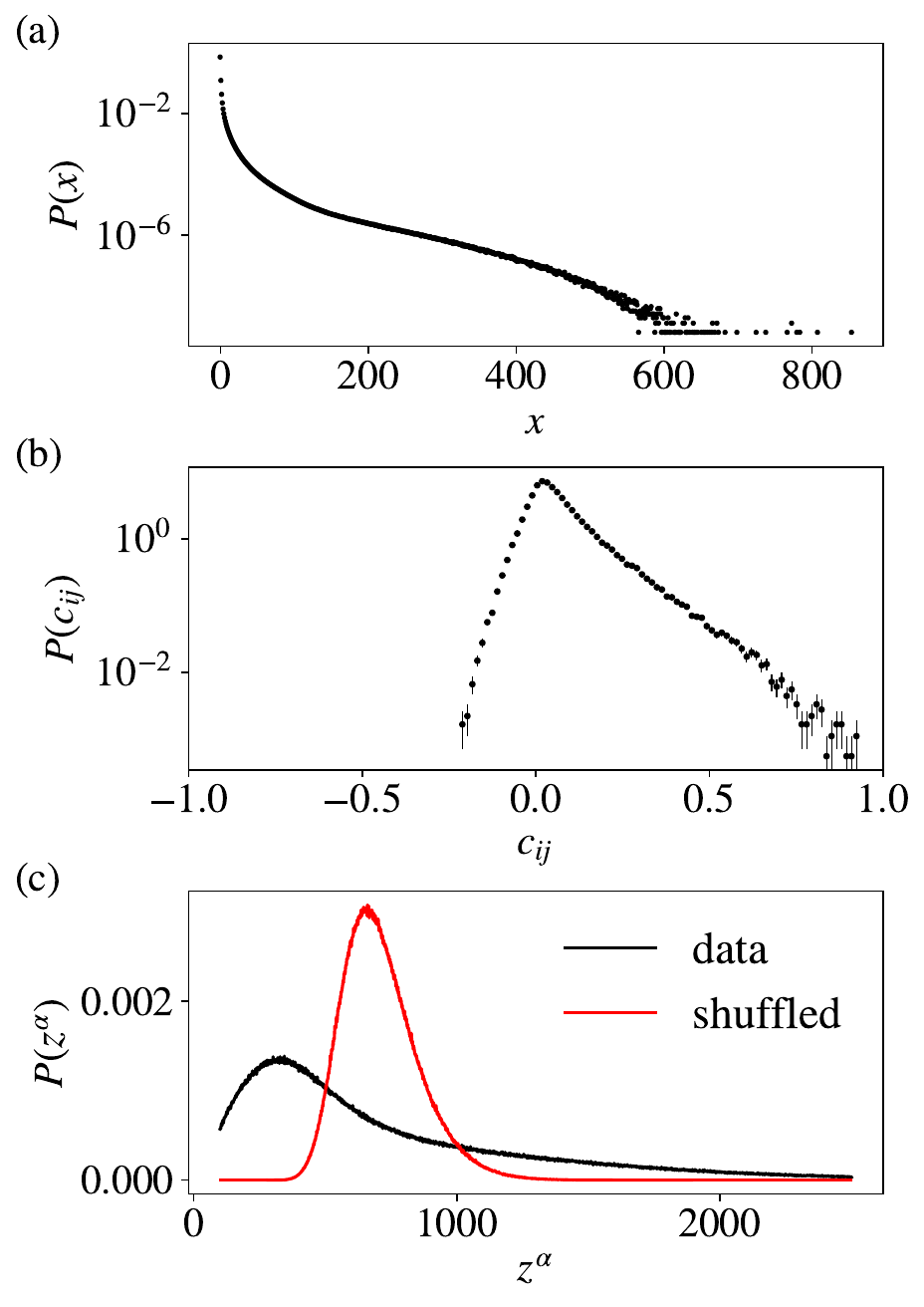}
    \caption{Basic features of the data.
        (a) Distribution of mRNA copy number for each gene, averaged over genes, $P(x)$ from \eqref{Px_def}.
        (b) Distribution of the correlation coefficient, $c_{ij}$ from \eqref{rhox}, across all $6.25\times 10^4$ pairs of genes.
        (c) Distribution of the summed expression level, $z^\alpha$ from \eqref{z_def}, in black for the real data and in red for shuffled data where all correlations are lost. This shows that interactions between mRNA species are not negligible.
        \label{basic_data}}
\end{figure}

In what follows, we focus on the MERFISH data.  For each gene $i = 1,\, 2,\, \ldots ,\, N_g = 500$, in the cell $\alpha$ there are $x_i^\alpha$ mRNA molecules; these counts are natural numbers. For each cell, we also know the class that was associated to it, $y^\alpha$.  We draw attention to several features of the data.  First, \figref{basic_data}a shows the distribution of expression levels averaged over genes,
\begin{equation}
    P(x) \equiv \langle P_{\rm i} (x_i = x)\rangle_i  = \frac{\sum_{i,\alpha} \delta_{x,x_i^\alpha}}{\sum_{i,\alpha} x_i^\alpha} .
    \label{Px_def}
\end{equation}
We see that this distribution has a long tail and a sharp peak at low expression levels; indeed fully $\sim 70\%$ of the weight is concentrated at $x=0$.  Second, the pairwise correlation coefficients between expression levels,
\begin{equation}
    c_{ij} = {\frac{\langle \delta x_i \delta x_j\rangle }{\sqrt{\langle (\delta x_i)^2\rangle\langle (\delta x_j)^2 \rangle}}},
    \label{rhox}
\end{equation}
where $\delta x = x - \langle x\rangle$, are largely positive.  A tail extends to correlations of almost unity, and since we have several million samples the threshold for statistical significance is $\rho \sim 10^{-3}$, essentially invisible in \figref{basic_data}b; this means there is no resolvable peak of uncorrelated pairs.  Finally, a corollary of these widespread correlations is that the distribution of summed expression levels,
\begin{equation}
    z^\alpha = \sum_i x_i^\alpha,
    \label{z_def}
\end{equation}
is very far from what we would observe if the genes were all expressed independently, as seen by comparing with shuffled data in \figref{basic_data}c.

\section{Cell classes, expression levels, and binary variables}

The classification of cells in the BICCN data is based on mRNA levels measured for many thousands of genes using scRNAseq.  The counting of molecules via MERFISH is more precise, but focuses only on 500 genes, so it is not clear that there will be enough information to determine even the major classes.

\begin{figure}[t]
    \centering
    \includegraphics[width=\linewidth]{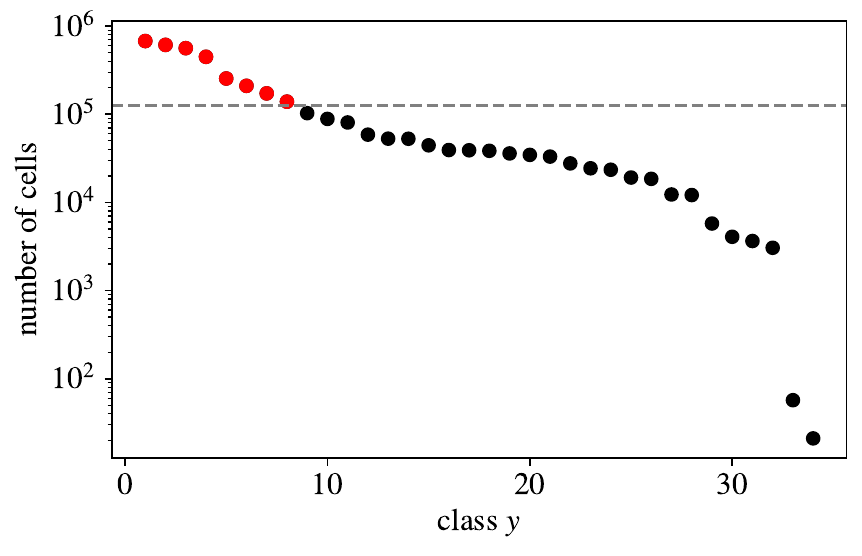}
    \caption{Selection of the 8 largest classes out of the total 34. Number of cells in each class, in rank order. The dashed line indicates the number of parameters of an Ising model with 500 variables. In the dataset used, only the first 8 classes have at least as many cells as parameters.}
    \label{fig:classes_size}
\end{figure}

To be sure that we do not run into sampling problems, we focus primarily on cells that belong to the eight largest classes out of the total of thirty--four (\figref{fig:classes_size}).  We choose eight classes because, looking ahead, we will build  models with pairwise interactions among genes, and this cutoff ensures that even within a single class we have enough samples to safely learn such complex models.
We then train a neural network \cite{goodfellow2016deep} to take as input the five hundred expression levels $\{x_i^\alpha\}$ for each cell $\alpha$ and return the label $y^\alpha = 1,\, 2,\, \ldots ,\, 8$ corresponding to the cell class.

\begin{figure}[b]
    \centering
    \includegraphics[width=\linewidth]{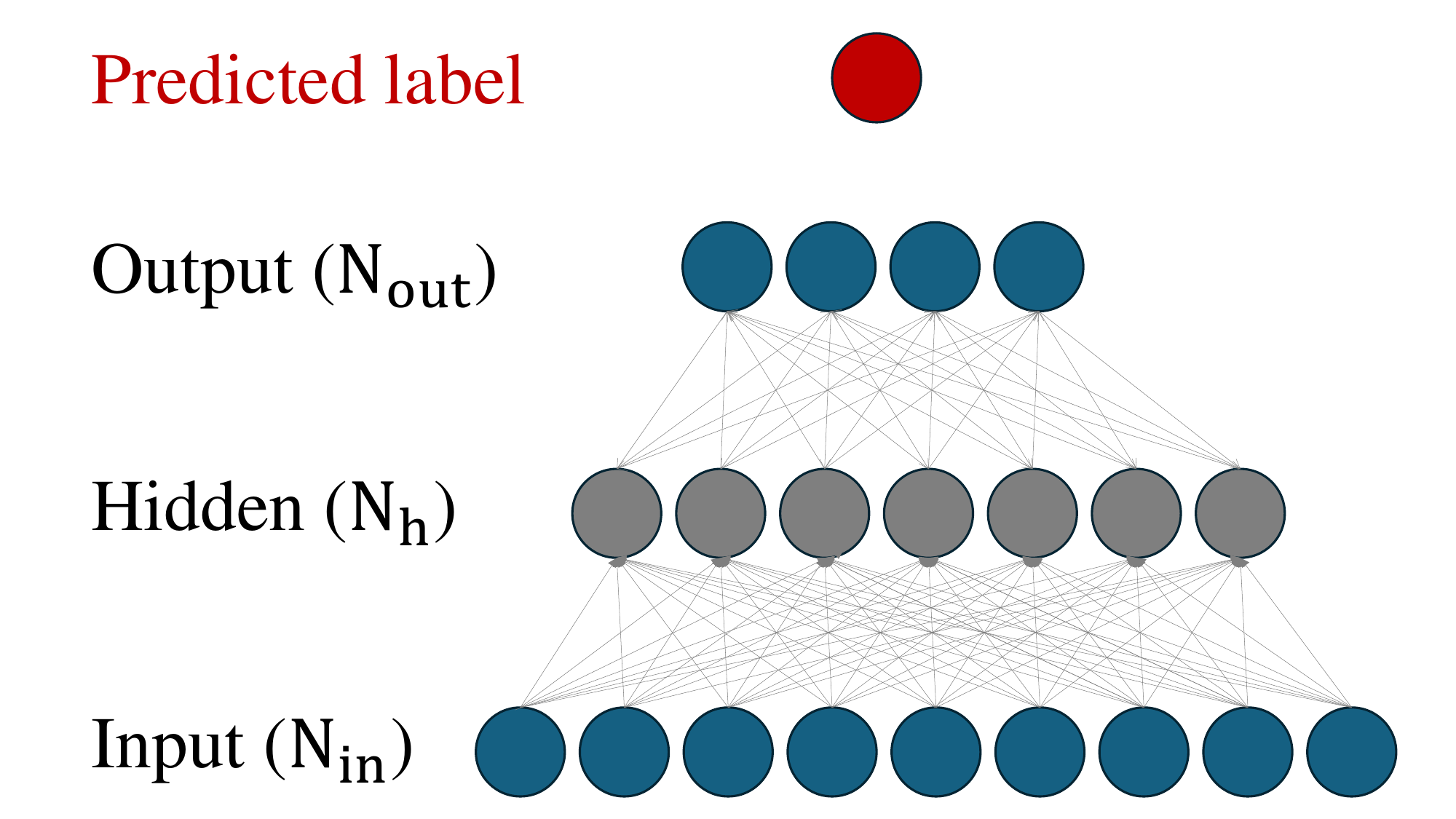}
    \caption{Schematic representation of the neural network classifier. Input layer with $N_{in}$ nodes (number of genes), hidden layer with $N_h = 100$ nodes, output layer with $N_{out} =8$  nodes (number of classes). The argmax over the output layer gives the predicted maximum likelihood label $\Tilde{y}$.}
    \label{fig:scheme_NN}
\end{figure}

The network, sketched in \figref{fig:scheme_NN}, is relatively simple: to classify a cell considering $N$ genes there are $N_{\text{in}} = N$ input nodes, $N_{h} = 100$ hidden nodes, and $N_{\text{out}} =8$ output nodes.
Each node is connected to all the nodes in the previous layer through the weights $\{W\}$ and the biases $\{b\}$, and the parameters are optimized by performing gradient descent on the cross entropy loss between predicted and known labels.
To be more specific, the outputs of the hidden nodes $\{\chi_j^\alpha\}_{j=1, \ldots, N_h=100}$
are given by
\begin{equation}
    \chi_j^\alpha = \text{ReLU}\left(\sum_{l=1}^{N_{\text{in}}} W_{l,j}^{(1)} x_l^\alpha + b_j^{(1)}\right) ,
\end{equation}
where $\text{ReLU}$ is the rectified linear unit, returning the positive part of the input:
\begin{equation}
    \text{ReLU}(x\leq0) = 0 \,\,\,\,\,\,\,\,\,\,\text{ReLU}(x>0) = x.
\end{equation}
    The value of the output nodes $\{k_c^\alpha\}_{c=1, \ldots, N_{\text{out}}}$ is given by
\begin{equation}
    k_c^\alpha = \sum_{j=1}^{N_h} W_{c,j}^{(2)} \chi_j^\alpha + b_c^{(2)}.
\end{equation}
To interpret it as a probability, the softmax of the output layer is considered, 
\begin{equation}
    \Tilde{p}_c^\alpha = \frac{\exp(k_c^\alpha)}{\sum_j \exp(k_j^\alpha)}.
\end{equation}
The weights and biases are updated so that $\Tilde{p}_c^\alpha$ is as close as possible to $\delta_{(c, y^\alpha)}$, where $y^\alpha$ is the true label.
This is done by defining a loss over all the cells in the batch
\begin{equation}
    L = -\frac{1}{N_{c}} \sum_{\alpha=1}^{N_{c}} \sum_{c=1}^{N_{out}} \delta_{(c, y^\alpha)} \log \Tilde{p_c}^\alpha
\end{equation}
and updating the parameters $\{W\}$,$\{b\}$ with gradient descent
\begin{eqnarray}
    &\Delta W= - \eta \nabla_W L\\
    &\Delta b = - \eta \nabla_b L
\end{eqnarray}
using Adam method~\cite{kingma2014adam} with a batch size of 64.

Finally the predicted label is the maximum likelihood label $\Tilde{y}^\alpha =\text{argmax}_c \Tilde{p}_c^\alpha$.

A random selection of 80$\%$ of the cells is used for training, the remaining $20\%$ is left for testing.
The performance of the classifier is measured by the accuracy, i.e. the fraction of test cells that are labeled correctly.  Note that one could always ignore the gene expression levels and assign cells to the largest class, which in this case will be correct a bit more than $20\%$ of the time; this sets a lower bound on performance.

\begin{figure}[t]
    \centering
    \includegraphics[width=1\linewidth]{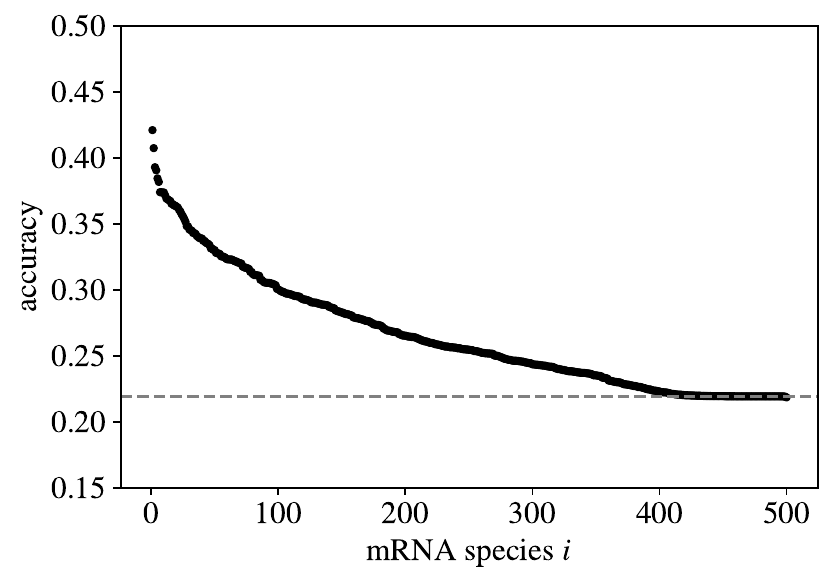}
    \caption{Cell classes are not determined by single genes.  We train the network in \figref{fig:scheme_NN} to take a single gene expression level $x_i$ as input and return the predicted cell class $\Tilde{y}$.  Performance is shown for all $N_g = 500$ choices of the single gene, in rank order.  The dotted line is the fraction of correct assignments that can be achieved naively by assigning all cells to the largest class.
        \label{singlegenes}}
\end{figure}

\begin{figure}[b]
    \includegraphics[width=\linewidth]{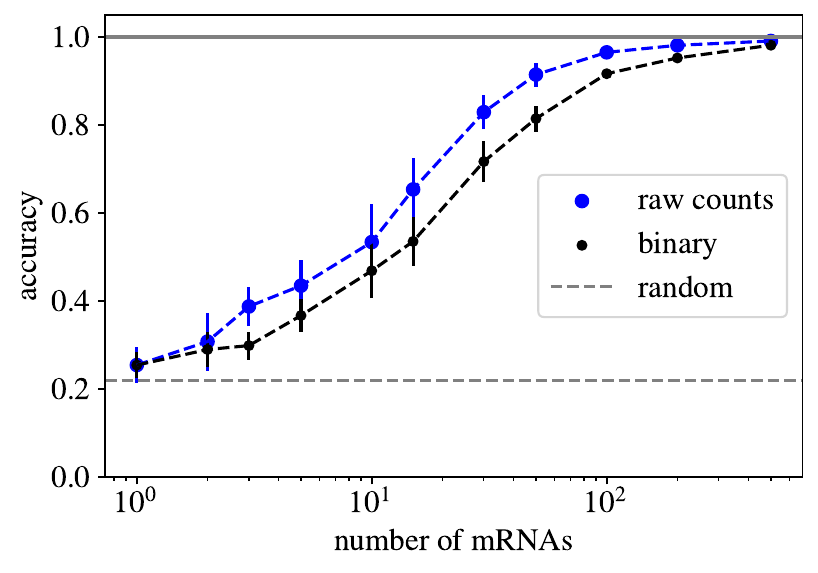}
    \caption{Classification accuracy. Average performance of the neural network trained to classify raw counts (in blue) and binary variables (in black) over 20 random sets of genes. The dotted line is the baseline obtained by assigning all the cells to the most likely class.
        \label{NNperformance}}
\end{figure}

It is important that single gene expression levels are not sufficient to solve the classification problem---that is, there are no ``markers'' for cell classes (\figref{singlegenes}).  On the other hand, if we choose $N$ genes at random out of the total of $N_g = 500$, then as we increase $N$ the performance improves steadily, reaching essentially perfect classification (\figref{NNperformance}).  Thus these 500 genes do indeed have enough information to define the cell class.

It is perhaps more surprising that we can substantially compress our description and still succeed in identifying cell classes.  As an example, suppose the we define a binary variable that tracks whether the expression level is zero or nonzero,
\begin{equation}
    \sigma_i^\alpha = \begin{cases}
        -1 & \text{if $x_i^\alpha =0$}    \\
        +1 & \text{if $x_i^\alpha \ne 0$}
    \end{cases}.
\end{equation}
We can retrain the neural network to use these binary variables as input, and, while there is some loss of information, we again see near perfect classification with sufficiently many genes (\figref{NNperformance}).

\begin{figure}[b]
    \centering
    \includegraphics[width=0.96\linewidth]{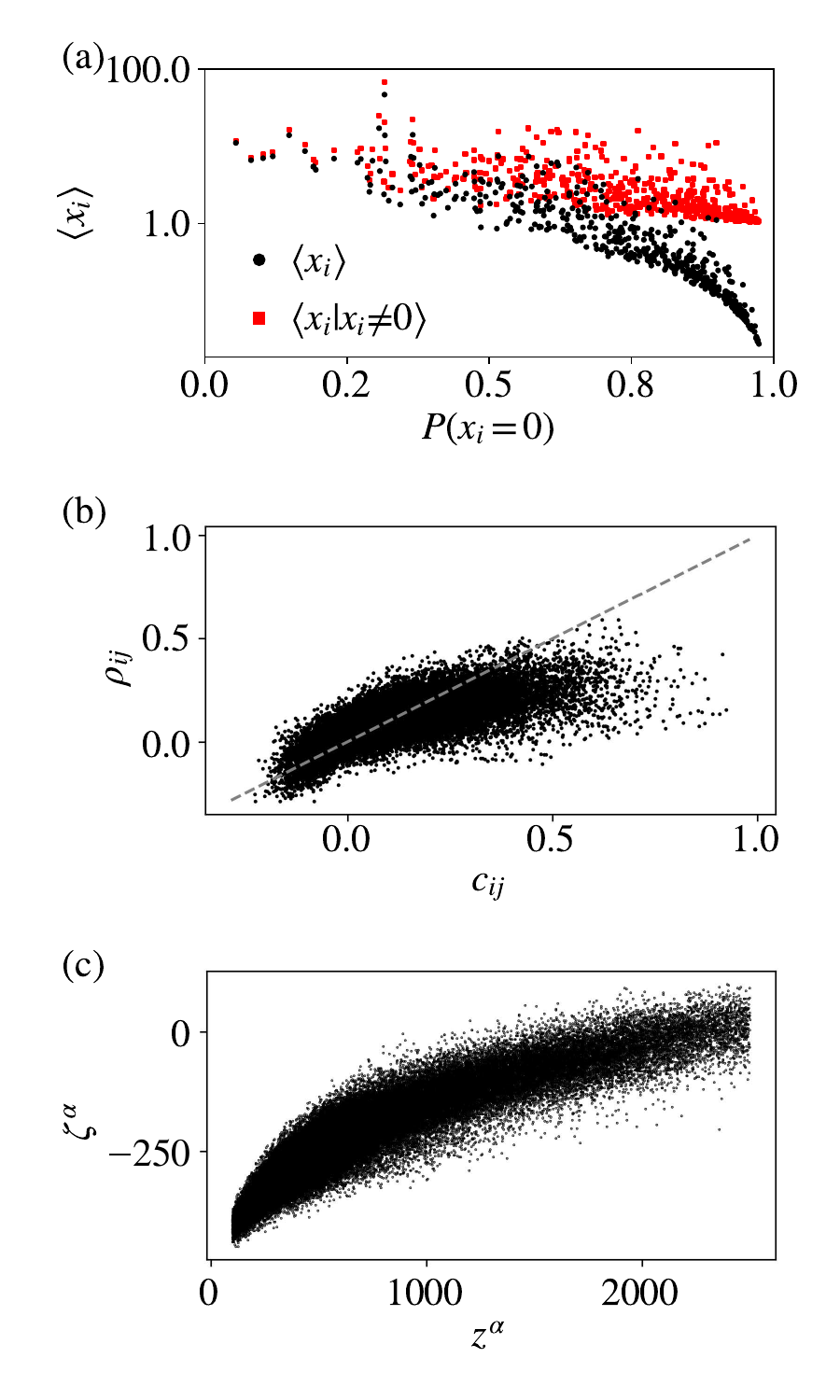}
    \caption{Data binarization.
        (a) Scatter plot of the zero-count probability vs the average counts. Each dot represents an mRNA species. The average is computed over all the cells (black) or only over the cells with at least one copy of that species (red).
        (b) Raw vs binarized variables correlation coefficients, from Eqs.~(\ref{rhosigma}) and (\ref{rhox}). Each dot represents a pair of mRNAs.
        (c) Comparison of total mRNA counts in a cell and the corresponding sum of binarized variables. Each dot represents a cell.}
    \label{fig:binarization}
\end{figure}

We can understand the success of the binary description by looking more carefully at the data.  We have seen that the distributions of individual gene expression can have a peak at $x=0$ that is distinguished from the bulk (\figref{basic_data}a).  This peak often contains more than half the weight of the distribution, and the value of $P_i(x_i = 0)$ is predictive of the mean nonzero value $\langle x_i | x_i \neq 0\rangle$ (\figref{fig:binarization}a).  Further, as shown in Fig   \ref{fig:binarization}b, the correlations between pairs of binary variables,
\begin{equation}
    \rho_{ij} = {\frac{\langle \delta \sigma_i \delta \sigma_j\rangle }{\sqrt{\langle (\delta \sigma_i)^2\rangle\langle (\delta \sigma_j)^2 \rangle}}},
    \label{rhosigma}
\end{equation}
are well predicted by the correlations of the corresponding raw variables, $c_{ij}$ from \eqref{rhox}.
Finally, the summed activity of all the genes $z^\alpha$  (\eqref{z_def}), whose distribution is shown in \figref{basic_data}c, is strongly correlated with the sum of the binary variables
\begin{equation}
    \zeta^\alpha = \sum_i \sigma_i^\alpha,
    \label{z_binary_def}
\end{equation}
in \figref{fig:binarization}c.

In summary, counting mRNA molecules for just 500 genes provides enough information to distinguish major cell classes, and much of the underlying statistical structure is captured by binary variables that distinguish zero vs nonzero expression.

\section{An Ising model for patterns of expression}

The fact that we can capture much of the structure of gene expression with binary variables leads us to ask if we can go further and write an effective approximation for the distribution over these variables, $P(\{\sigma_i\})$.  Following work on patterns of activity in networks of neurons ~\cite{schneidman2006weak,tkavcik2014searching,granot2013stimulus,meshulam2017collective}, ordering of flight velocities in flocks of birds ~\cite{bialek2012statistical,mora+al_16}, and the evolution of sequences in protein families ~\cite{bialek+ranganathan_07,weigt+al_09,mora2010maximum,marks2011protein,morcos2011direct,bitbol2016inferring}, we use the maximum entropy method ~\cite{jaynes1982rationale}.

\subsection{Building the model}

In the maximum entropy method we look for the least structured probability distribution that is consistent with a set of measured expectation values.  Quite generally, if we define important observables $f_\mu (\{\sigma_i\})$, then we insist that our model for the probability distribution $P(\{\sigma_i\})$ predict the expectation values of these observables correctly, so that
\begin{equation}
    \langle f_\mu(\{\sigma_i\})\rangle_P = \langle f_\mu(\{\sigma_i\})\rangle_{\text{expt}} .
    \label{constraints}
\end{equation}
Then the maximum entropy distribution that satisfies these constraints has the form
\begin{eqnarray}
    P (\{{\sigma_i}\}) &=& {\frac{1}{Z}} \exp \left[- E(\{\sigma_i\})\right]\label{me1}\\
    E(\{\sigma_i\}) &=&  \sum_\mu g_\mu  f_\mu(\{\sigma_i\})  ,  \label{me2}
\end{eqnarray}
where the couplings $g_\mu$ have to be adjusted to satisfy \eqref{constraints} for all the constraints $\mu$.

In our case, we want to match the mean expression level of each gene, or more precisely the probability of nonzero expression $p_i = ( 1 + \langle \sigma_i \rangle )/2$.  If these were the only expectation values that we match, then the maximum entropy model would describe each gene turning on and off independently.  As a first step to describing interactions and hopefully collective behavior in the network we will also match the correlations between pairs of genes,  $\langle\sigma_i\sigma_j\rangle$.  With these choices for the observables $\{f_\mu\}$,  the maximum entropy distribution takes the form of an Ising model, where \eqref{me2} is replaced by the more explicit
\begin{equation}
    E(\{\sigma_i\}) =  \sum_i h_i \sigma_i + {\frac{1}{2}} \sum_{i> j} J_{ij} \sigma_i \sigma_j .
    \label{ising1}
\end{equation}
Again, the ``fields'' $\{h_i\}$ and ``couplings'' $\{J_{ij}\}$ need to be adjusted so that the predictions of the model match the measured expectation values $\{\langle\sigma_i\rangle\}$ and $\{\langle\sigma_i\sigma_j\rangle\}$.

\begin{figure}[t]
    \centering    \includegraphics[width=1\linewidth]{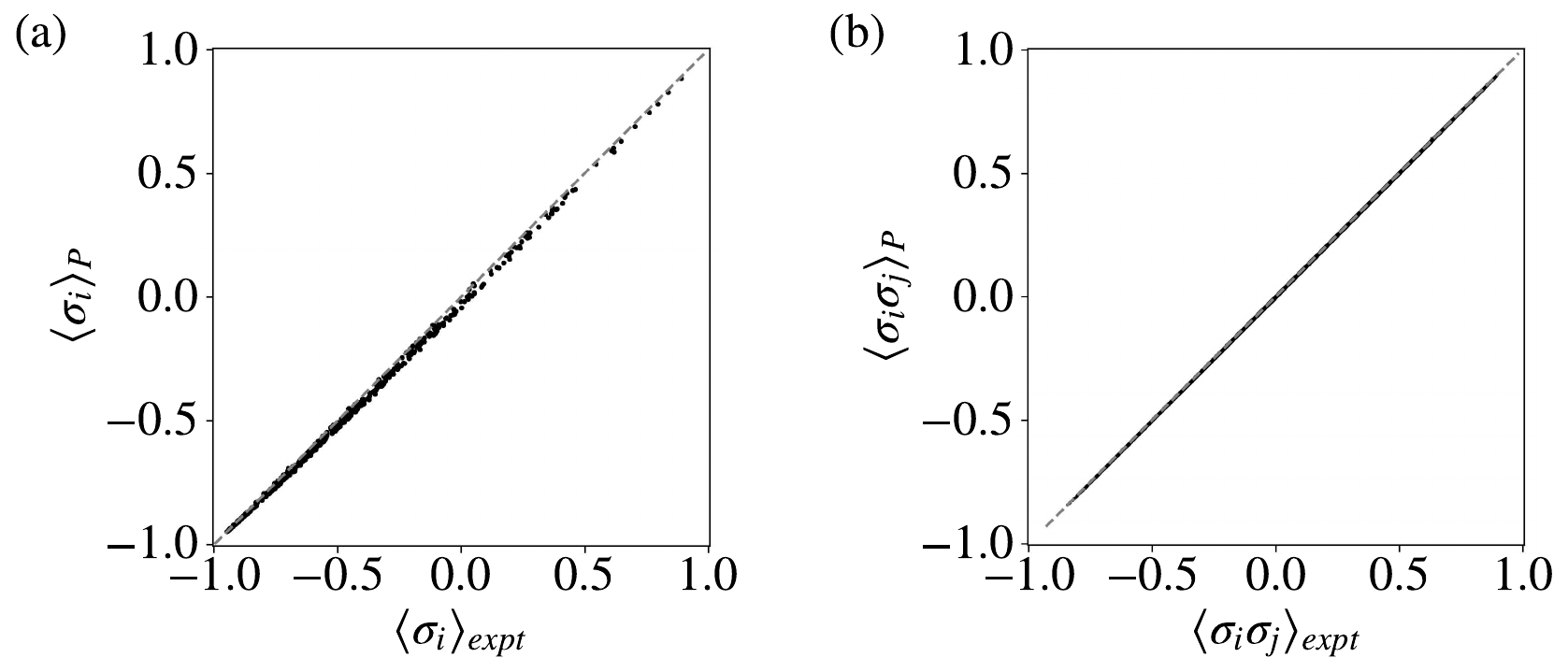}
    \caption{Convergence of constraints for the Ising model with all the $N=N_g = 500$ mRNA species. Comparison of first (a) and second moments (b).
    }
    \label{fig:ising_convergence}
\end{figure}

The ``inverse Ising'' problem of finding $\{h_i; J_{ij}\}$ to match $\{\langle \sigma_i \rangle , \langle \sigma_i \sigma_j \rangle \}$ is well known, and we follow standard methods   ~\cite{tkacik2006ising}.    Briefly, given a particular set of parameters we do a Monte Carlo simulation of the corresponding model in Eqs.~(\ref{me1}, \ref{me2}) to estimate the expectation values $\langle f_\mu \rangle_P$.  Then we adjust all the parameters $g_\mu \rightarrow g_\mu + \Delta g_\mu$, where
\begin{equation}
    \Delta g_\mu = \eta \left(\langle f_\mu \rangle_{\rm expt} -\langle f_\mu \rangle_P\right) .
\end{equation}
The learning rate $\eta$ was varied during the training, also depending on the size of the system, and ranged from $10^{-2}$ to $10^{-5}$. The parameters of the independent model with the same means were used for the initialization. \figref{fig:ising_convergence} shows that we can reach good convergence, satisfying the constraints even for a large system with $N=500$~\footnote{As far as we know, the largest Ising models for neuronal activity have been with $N\sim 150$ cells ~\cite{tkavcik2014searching}, so reaching $N=500$ shows that significantly larger systems can be treated by this method.  The obstacle to maximum entropy models for larger numbers of neurons is not conceptual or computational, but simply that data sets have not been large enough to determine all $\sim N^2/2$ pairwise correlations reliably. To have a number of (reasonably independent) samples comparable to the MERFISH data used here, neural recordings would have to be $\sim 100\,{\rm hrs}$ long.}.

\subsection{Testing the model}

The pairwise Ising model in \eqref{ising1} is well motivated but of course not guaranteed to be correct.  Here we focus on testing the model for all $N=N_g = 500$ genes; see Appendix \ref{app:extra plots} for examples with $N=100$ and $N=200$.

\begin{figure}[b]
    \centering
    \includegraphics[width=1\linewidth]{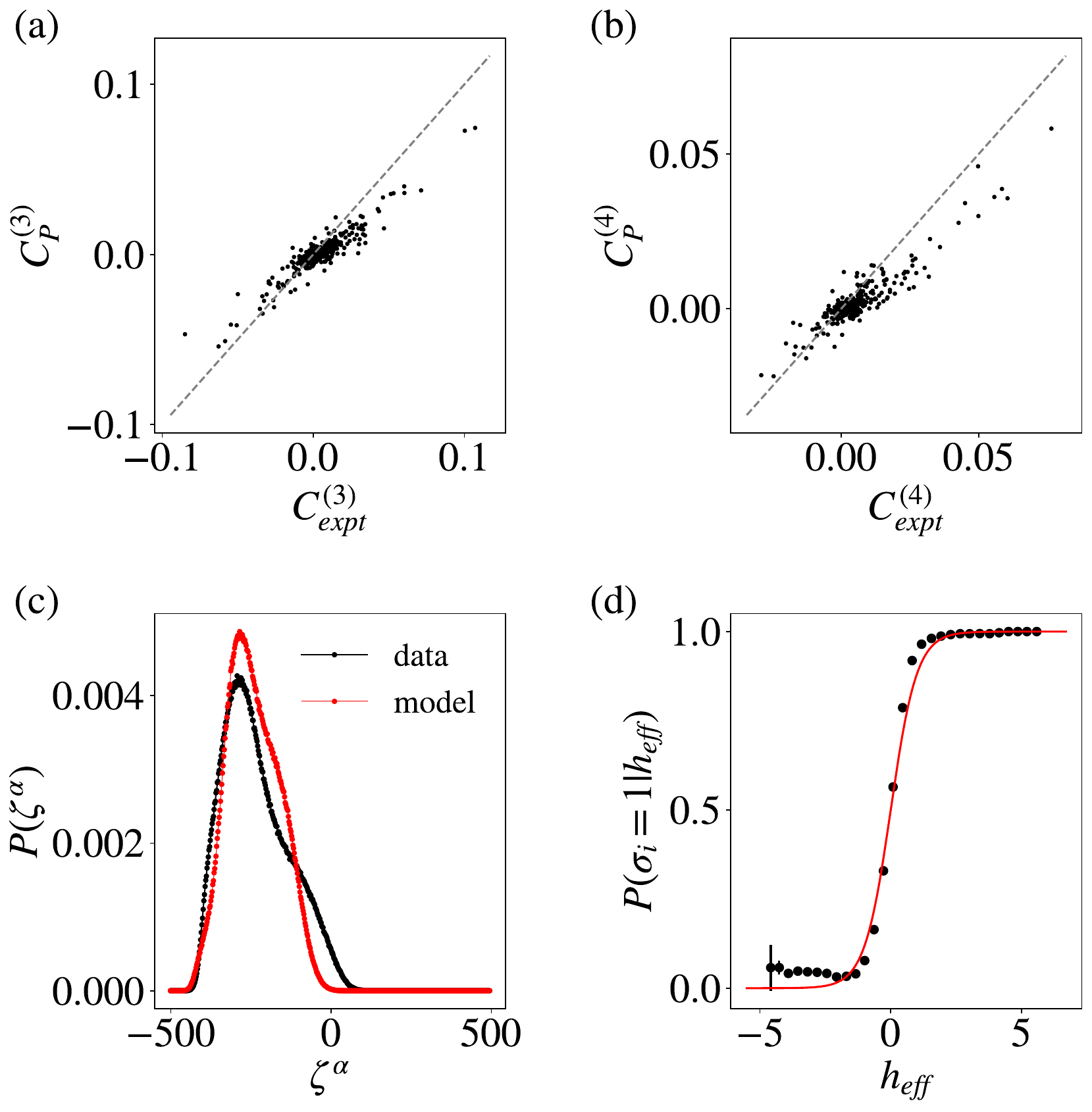}
    \caption{Testing the Ising model with all the $N=500$ genes.
        (a) Third-order connected moments. Comparison between data moments and Ising model predictions for 500 random combinations. The diagonal line represents perfect agreement between data and model predictions.
        (b) Fourth-order connected moments for 500 random combinations.
        (c) Sum distribution (\eqref{z_binary_def}): comparison between data (black) and Ising model (red).
        (d) Probability of one variable given the state of all the other genes, averaged over all the genes.  The red curve is given by \eqref{eq:heff}.
        \label{fig:ising_higher}}
\end{figure}

Since the Ising model is constructed from pairwise correlations, the first natural test is to check the predictions for higher-order correlations
\begin{eqnarray}
    C^{(3)}_{ijk} &=& \langle \sigma_i \sigma_j \sigma_k\rangle_c\\
    C^{(4)}_{ijkl} &=& \langle \sigma_i \sigma_j \sigma_k \sigma_l \rangle_c,
\end{eqnarray}
where $\langle \cdots \rangle_c$ denotes the connected part. In Figures~\ref{fig:ising_higher}a and  \ref{fig:ising_higher}b we see that these predictions agree reasonably well with experimental data at $N=500$, although these values are small and there is some scatter.

A more global test of the model's behavior is to look at the distribution of the summed activity $\zeta^\alpha$  \eqref{z_binary_def} which we have seen is very far from what would happen if the genes were turned on and off independently (\figref{basic_data}c).  Again the agreement between theory and experiment is quite good, although the predicted $P(\zeta^\alpha)$ is a bit narrower than seen in the data (\figref{fig:ising_higher}c); this is consistent with the slight underestimate of higher moments seen in Figures~\ref{fig:ising_higher}a and b.

Finally, we can interpret our model for the {\em joint} distribution of gene expression levels as a model for the conditional level of expression of one gene given the state of all the other genes.  Concretely, \eqref{ising1} predicts that
\begin{eqnarray}\label{eq:heff}
    P(\sigma_i =1| \{\sigma_{j\neq i}\}) &=& {\frac{1}{1+ \exp[-2 h_{\rm eff}(\{\sigma_{j\neq i}\})]}}\label{PvsH}\\
    h_{\rm eff}(\{\sigma_{j\neq i}\}) &=& h_i + \sum_{j\neq i} J_{ij}\sigma_j .
\end{eqnarray}
We can test this prediction by sifting through all cells and all individual genes, evaluating $h_{\rm eff}$, then making small bins along the $h_{\rm eff}$ axis and estimating the probability that $\sigma_i = 1$ for cells and genes that fall into each bin.
Results are shown in \figref{fig:ising_higher}d, compared with the simpler prediction of \eqref{PvsH}; again the agreement is quite good.

\begin{figure}[t]
    \centering
    \includegraphics[width=\linewidth]{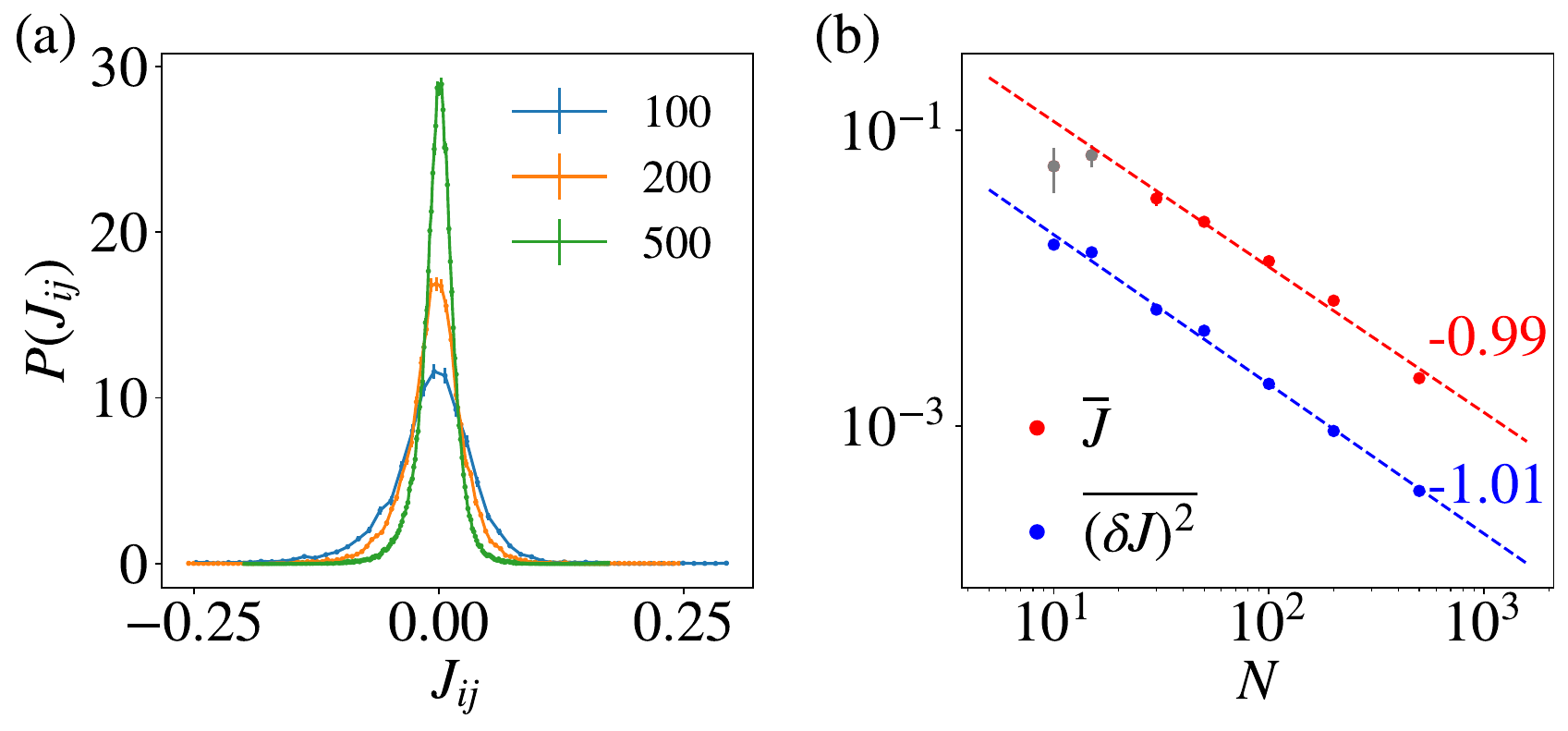}
    \caption{Distribution of the interactions for different system sizes.
        (a) Even if the correlations are almost all positive as shown in \figref{basic_data}b, $J$ have both positive and negative values.
        (b) Scaling behavior of mean and variance of couplings. The fit of the mean value is computed excluding the first two points.
        \label{fig:j}}
\end{figure}

\subsection{Thermodynamics?}
We can build the model in \eqref{ising1} for different numbers of genes $N$. We select the set of genes for each $N$ from 150 random choices as the one that gives the best classifier performance using the neural network classifier.
While we find that the fields do not vary systematically with $N$, on the other hand, we see in \figref{fig:j} that there is a systematic variation in the mean coupling
\begin{equation}
    \bar J \equiv {\frac{1}{N(N-1)}} \sum_{i \neq j} J_{ij} \sim 1/N
\end{equation}
and the variance of couplings
\begin{equation}
    \overline{(\delta J)^2} \equiv {\frac{1}{N(N-1)}} \sum_{i \neq j} (J_{ij} - \bar J)^2 \sim 1/N .
\end{equation}
These are the behaviors required for the existence of a thermodynamic limit in a broad class of models ~\cite{mezard1987spin}, but here they emerge from the data.  This  suggests that we are seeing signs of $N\rightarrow\infty$ behavior once we look at $100+$ genes.

Since the models we consider here are equivalent to equilibrium statistical mechanics problems, it is natural to ask if there is a corresponding thermodynamics at large $N$.  We have seen that this is possible because of the scaling of the interactions, but it is not obvious what this thermodynamics would look like or more specifically where in the phase diagram of possible models we will find the model that describes the real system.

\begin{figure}[t]
    \centering
    \includegraphics[width=1\linewidth]{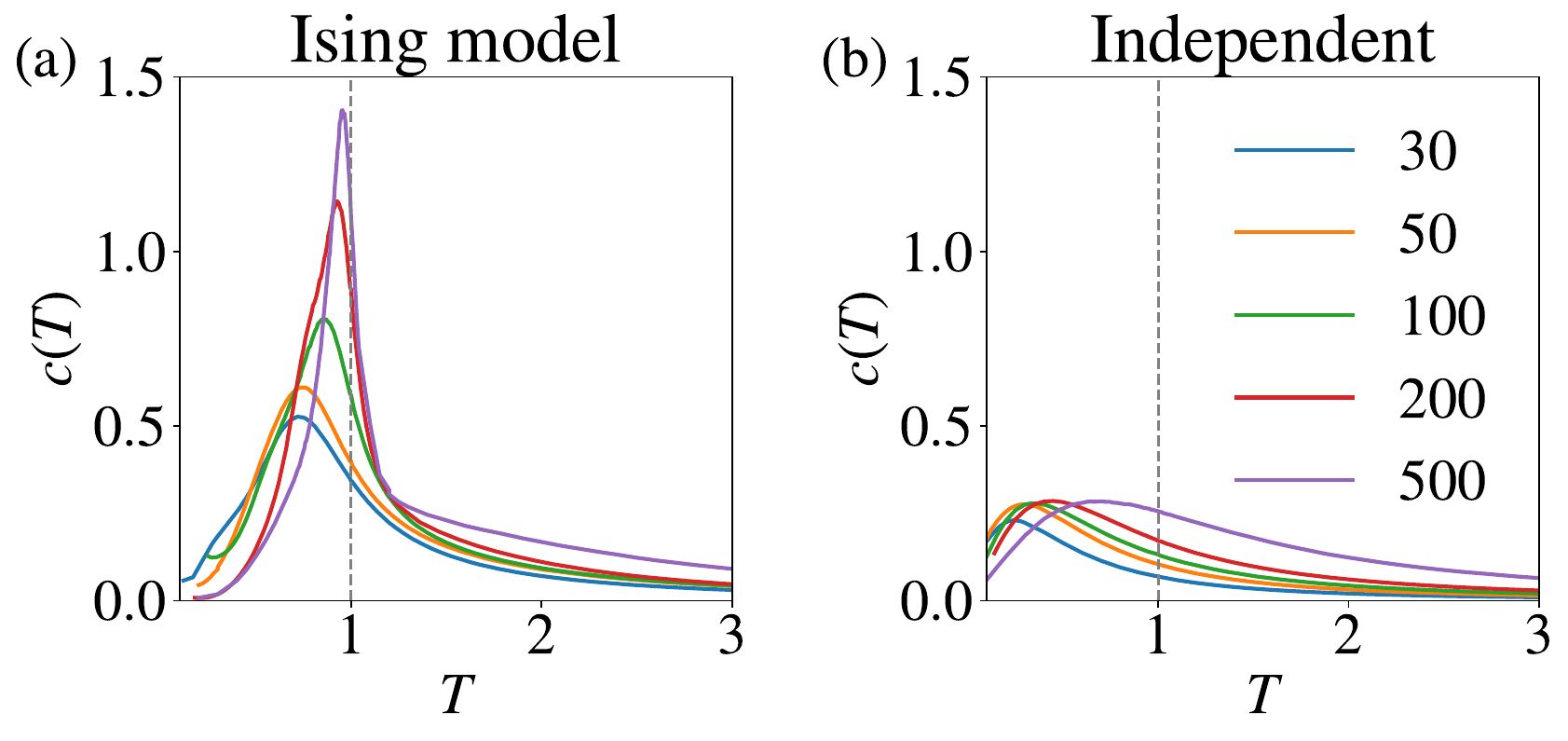}
    \caption{Specific heat, from \eqref{Cvdef}. (a) For Ising models, as in Eqs~(\ref{me1}, \ref{ising1}), constructed with varying numbers of genes. (b) For models of genes independently turning on and off.
        \label{fig:hc_s}}
\end{figure}

To explore the space of models, we can introduce in \eqref{me1} a fictitious temperature parameter $T=1$ for real data and study what happens as we vary this temperature ~\cite{tkacik2014thermodynamics}.
This corresponds to scaling the interactions and fields (collectively denoted with $\{ g_\mu\}$) as
\begin{equation}
    g_\mu \to \frac{g_\mu}{T} ,
\end{equation}
which allows us to test if the model describing real data ($T=1$) is special in some way.
At each temperature $T$ and for each size of the system $N$ we can compute the specific heat as
\begin{equation}
    c(T) = \frac{\langle (\delta E)^2 \rangle}{T^2 N}.
    \label{Cvdef}
\end{equation}
In \figref{fig:hc_s}a we see that the peak in the specific heat becomes higher, sharper, and closer to $T=1$ as we consider larger $N$.
On the contrary, when we consider a model in which genes are turned on and off independently the specific heat has a broad peak well below $T=1$ for all $N$, and the maximum $c(T)$ is independent of $N$ (\figref{fig:hc_s}b).

\begin{figure}[b]
    \centerline{\includegraphics[width =0.8\linewidth]{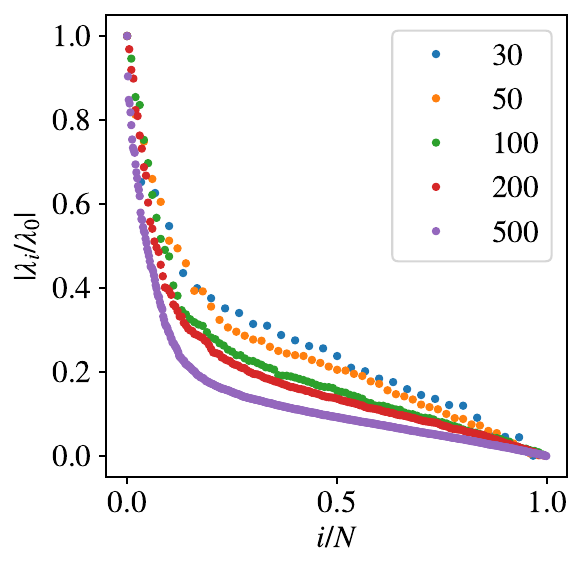}}
    \caption{Interactions are of full rank.  Eigenvalues of the matrix $J_{ij}$ in the Ising models that describe particular combinations of $N=30, 50, 100,$ and $200$ genes, as well as the full network of $N=500$ genes.  Eigenvalues are normalized by their largest value and plotted vs fractional rank. \label{eigJ}}
\end{figure}

The behavior of the specific heat that we see in \figref{fig:hc_s}a reminds us of that near a critical point, although we should be cautious in our interpretation.  The growth of the specific heat with $N$ is a sign that correlations extend throughout the network, even though each gene almost certainly interacts directly with only a small number of others.  One could achieve the same effect by having all genes coupled to some hidden variable(s)\footnote{Recall that any model of interacting spins is equivalent to a model of independent spins interacting with correlated fluctuating fields.  The question is whether this rewriting of the model constitutes a simplification, or points to physically meaningful hidden degrees of freedom.}, but if there are only a small number of these variables then we might expect the effective interactions to be of low rank, and this emphatically is not the case (\figref{eigJ}).  Note that since we learn our models from $\sim 10^6$ samples, the full spectrum of interactions is significant.  If we imagine a network with the same mean expression levels $\langle \sigma_i\rangle$ but uniformly weaker correlations, then the peak in the heat capacity will decrease, broaden, and move further below $T=1$, eventually approaching the independent model in \figref{fig:hc_s}b.  We conclude that the observed correlations indeed are in a special range, driving the models that we learn toward a critical point in their phase diagram.

\subsection{Local minima and cell classes?}

Although the correlations between gene expression levels are largely positive, the  interactions $J_{ij}$ in our model have both signs, allowing for competition and frustration.  We therefore expect, and find, that the energy function $E(\{\sigma_i\})$ in \eqref{ising1} has multiple local minima, states where the energy cannot be lowered by flipping any single binary variable $\sigma_i \rightarrow -\sigma_i$.

\begin{figure}[t]
    \centering
    \includegraphics[width=1\linewidth]{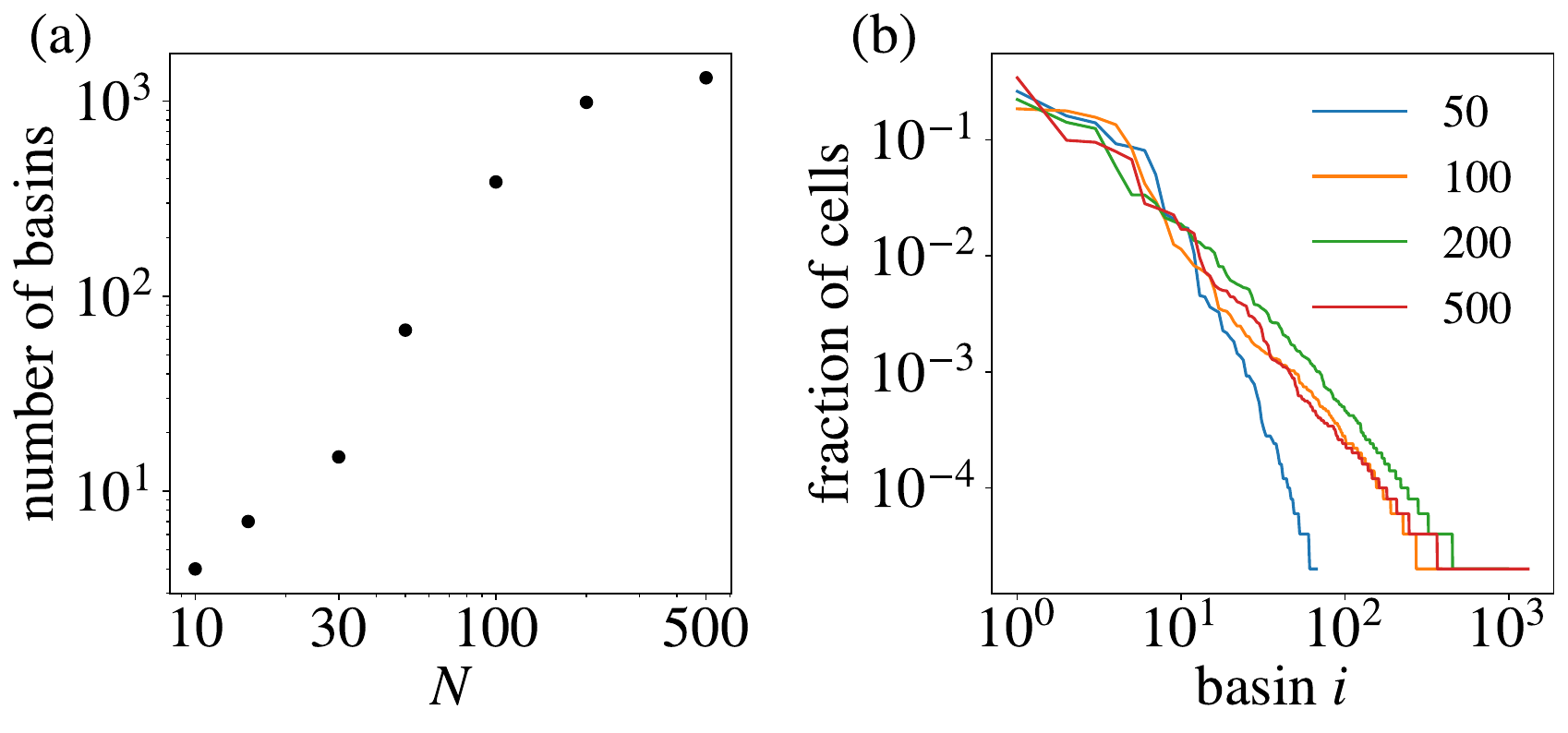}
    \caption{Energy landscape has multiple local minima.
        (a) Number of local minima for systems of different sizes $N$.
        (b) Fraction of cells in each energy basin. Different colors represent different sizes of the system.
        \label{fig:basins_sizes}}
\end{figure}

We find local minima of the energy by starting from actual states observed in the data and then moving ``downhill'' in a zero temperature Monte Carlo with spins flipped one at a time.   The result is that models built for larger numbers of genes have more local minima, up to $\sim 10^3$ when we include all $N= 500$ genes (\figref{fig:basins_sizes}a).

We characterize the size of the basins surrounding each local minimum by counting the fraction of observed states that relax to that minimum under the downhill dynamics.  The basins vary in size by orders of magnitude, and at large $N$ we see an approximately power--law tail to the distribution of sizes (\figref{fig:basins_sizes}b).

\begin{figure}[b]
    \centering
    \includegraphics[width=1\linewidth]{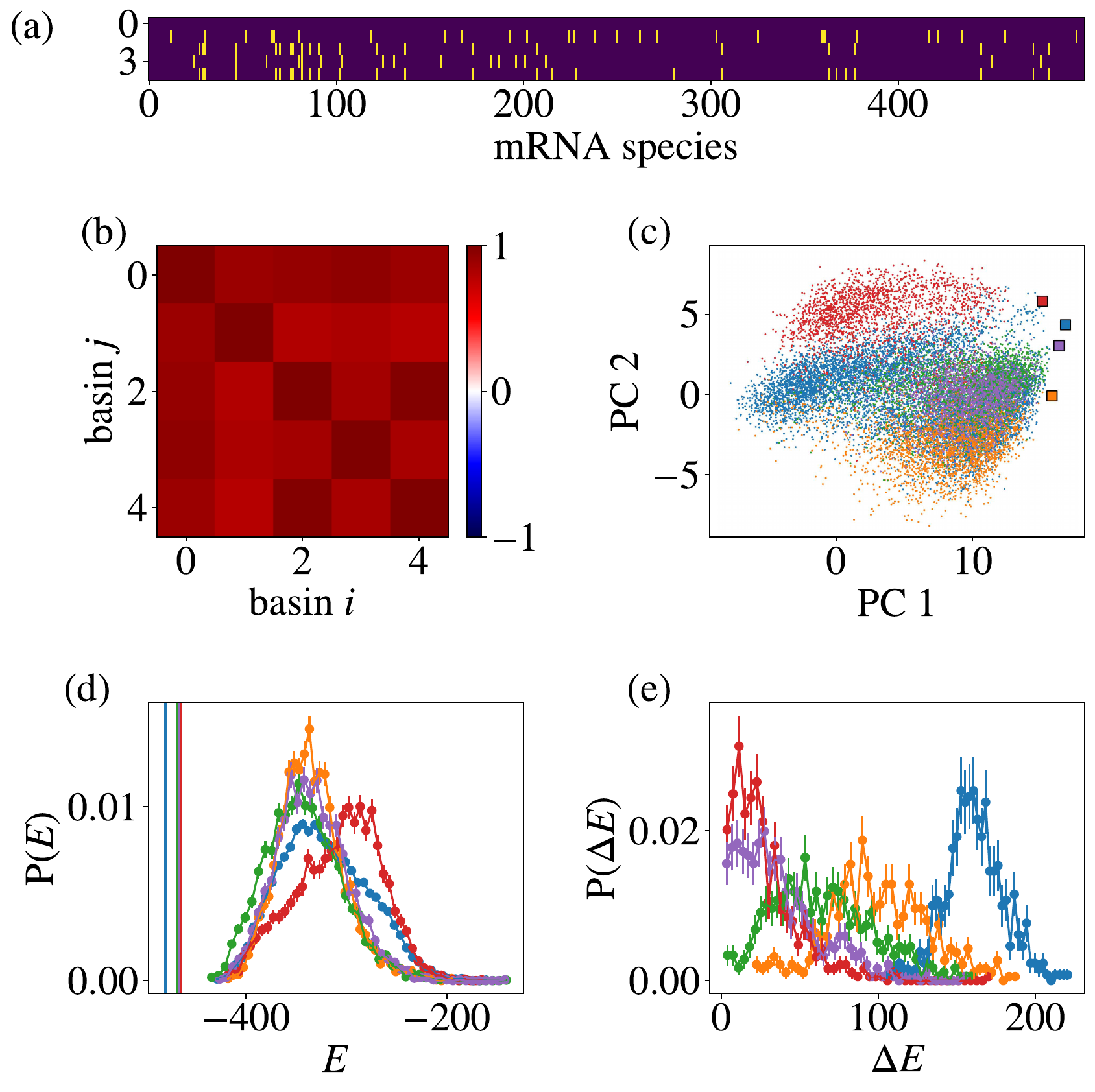}
    \caption{Energy basins for $N=500$.
        (a) Local minima $\{\sigma_i^A\}$ corresponding to the five largest basins. Each row is a different minimum, with columns being mRNA species.
        (b) Overlap between the first five minima $O_{AB} \equiv \sum_i^N \frac{\sigma_i^A \sigma_i^B}{N}$. Minima are quite similar to each other and far from orthogonal.
        (c) Visualization of the five largest basins (small dots) and the corresponding minima (squares) in the plane of the first two principal components of the data, capturing $\sim 13 \%$ of the total variance. Cells that belong to the same basin tend to cluster together.
        (d) Distribution of the energy of cells belonging to the same basin, for the five largest basins. Each color corresponds to a basin. Vertical lines represent the energy in the minima.
        (e) Distribution of energy barriers between minima, defined as the maximum energy difference over 500 trajectories that start from the minimum and end up in a different basin.}
    \label{fig:ising_basins}
\end{figure}

In systems such as the Hopfield model for associative memory, different local minima of the energy correspond to orthogonal configurations of the microscopic variables ~\cite{hopfield1982neural, anderson1995introduction}.  The mean--field spin glass is in an opposite limit, where the overlaps between local minima have a complex, hierarchical structure ~\cite{mezard1987spin}.  We see that in the network of genes, even the dominant minima (\figref{fig:ising_basins}a) are not orthogonal (\figref{fig:ising_basins}b), and overlaps between minima are large.

To visualize the local minima we can perform a principal components analysis of the states $\{\sigma_i\}$ and project onto the two components that carry the most variance, as shown in \figref{fig:ising_basins}c.  We see that the different basins of attraction are reasonably (though not completely) separable even in this low dimensional projection, and that the cells within a basin do not significantly populate the closest local minimum.  This makes sense because the fluctuations are high dimensional and have a significant entropy, even if we condition on the basin.

While the energies of cells in different basins are quite similar to each other (\figref{fig:ising_basins}d), the energy barriers to change basins are large, as seen in \figref{fig:ising_basins}e.
Energy barriers can be computed by starting from one of the minima and simulating a Monte Carlo trajectory until the state belongs to a different energy basin. The difference between the highest energy reached along the trajectory and the original energy gives an estimate of the barrier to overcome to leave the state. This is repeated 500 times for each minimum to get a distribution of barrier heights.

In summary, the energy landscape of our model consists of multiple local minima separated by reasonably large barriers.  Leaving the language of energies, we can say that the probability distribution is predicted to have multiple local peaks, separated by deep troughs.  It is tempting to associate the peaks with distinct classes of cells.  Could these correspond to the classes of cells identified experimentally?

\begin{figure}
    \centering
    \includegraphics[width=1\linewidth]{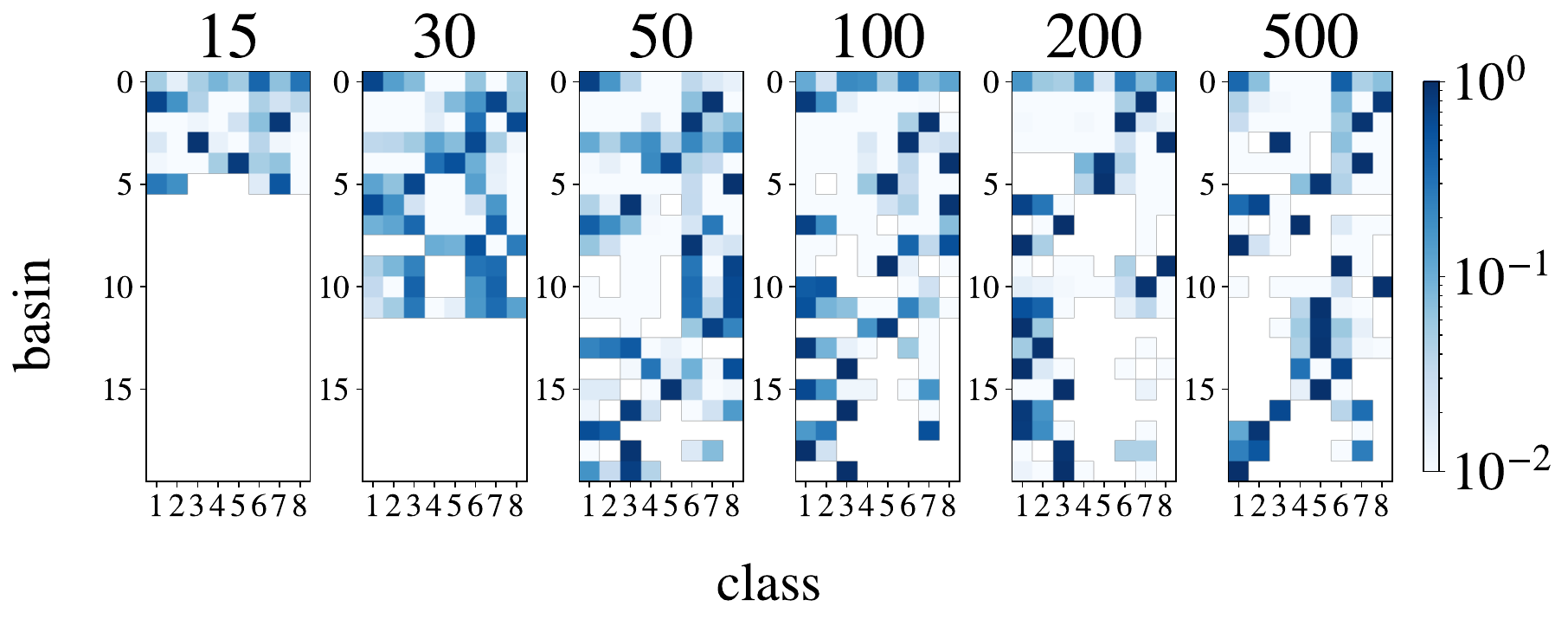}
    \caption{Composition of the 20 largest energy basins for different system sizes. Rows represents basins, sorted from the largest, columns are known classes. Intensity of color is the probability of a cell in a given basin to belong to a certain class, so sum over rows is 1. For small systems there are less than 20 basins, so the lowest rows are blank. Basins tend to be constituted of cells from the same class, suggesting a correspondence between energy basins and cell types.
        \label{fig:entropies_1ising}}
\end{figure}

\begin{figure}[b]
    \centering
    \includegraphics[width=1\linewidth]{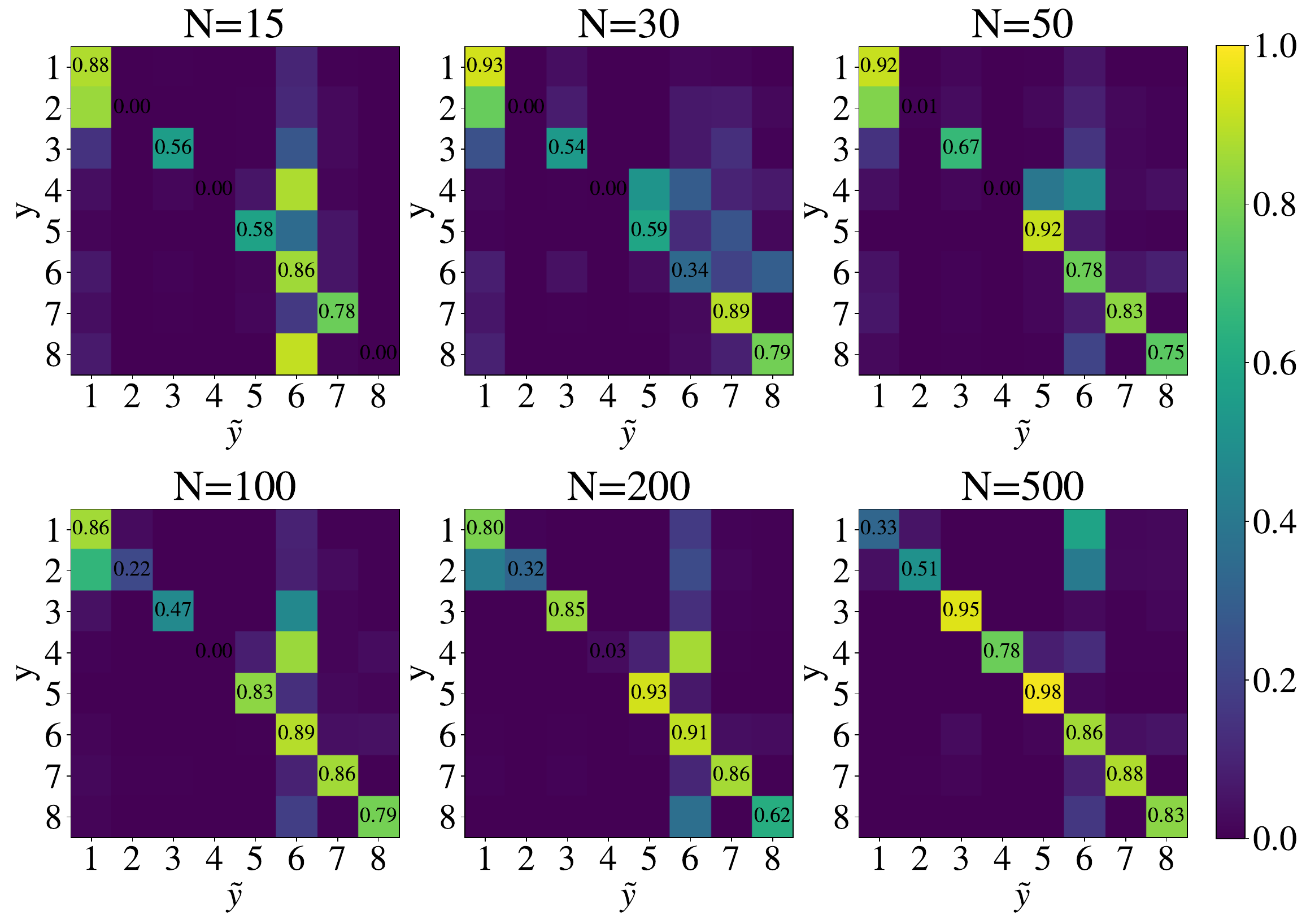}
    \caption{Confusion matrix for the Ising classifier for different sizes of the system (one choice of genes per each size). The color represents the value of $P(\tilde{y}|y)$, and the sum across each row is normalized to 1. The diagonal shows the probability that cells in class $y$ are correctly assigned to that class.
        \label{fig:confusion_binary}}
\end{figure}

We can map basins to cell classes by maximum likelihood by asking for the class to which most cells in the basin are assigned experimentally (\figref{fig:entropies_1ising}).
Most basins are made only or mainly of one kind of cells, suggesting a correspondence with cell types.
This provides a deterministic map from states to basins and then to classes, which we can summarize as $\{\sigma_i \} \rightarrow \tilde y$.
We can test the accuracy of this map by estimating the probability that cells that have been labelled as belonging to class $y$ are assigned to class $\tilde y$, $P(\tilde y | y)$, sometimes called the confusion matrix; results are shown in \figref{fig:confusion_binary}.  At small $N$ confusions are significant, but at $N=N_g = 500$ the energy basins of the Ising model map to cell classes more precisely, reaching  $\sim 90\%$ accuracy for several of the classes.

\begin{figure}
    \centering
    \includegraphics[width=\linewidth]{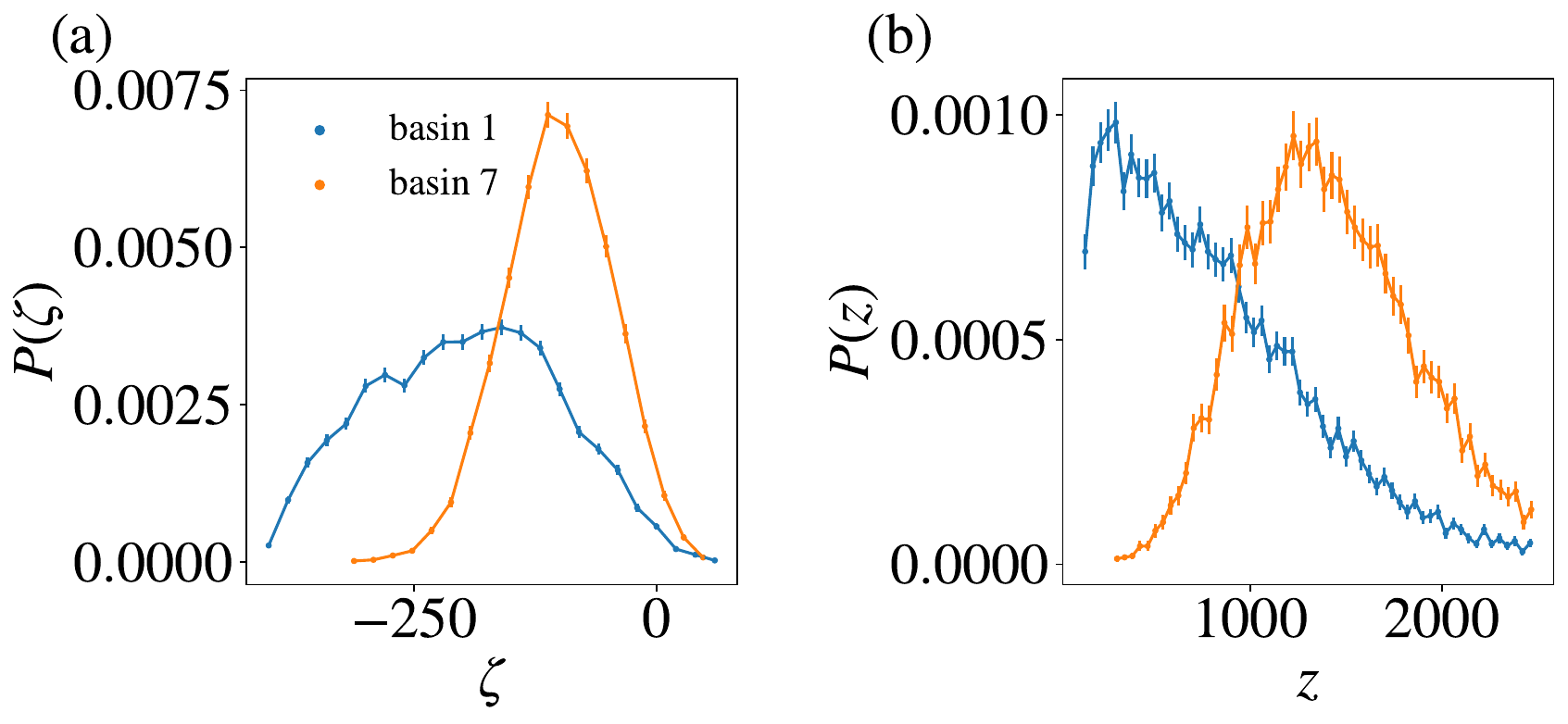}
    \caption{Cells of class $y=2$ fall into two basins.  (a) The ``magnetization'' $\zeta$ from \eqref{z_binary_def} has a very different distribution for cells in the two basins. (b) The summed value of raw mRNA counts, $z$ from \eqref{z_def} are similarly distinguishable.
        \label{fig:class2}}
\end{figure}

We can go a bit further and ask if the cases that we get ``wrong'' might be more subtle.  As an example, cells that have been assigned to class $y = 2$ in the original analysis of the data are split almost evenly between two basins in our energy landscape.  If we look just at these cells, the ones that fall in different basins are very distinguishable, even if we take just the one--dimensional ``magnetization'' $\zeta$ as a measure, as shown in \figref{fig:class2}a.  We can check that this is not an artifact of binarization by looking at the total mRNA count $z$, and we see that these counts also come from distinguishable distributions (\figref{fig:class2}b).

If energy basins---or, equivalently, peaks in the joint probability distribution of expression levels---have a mapping to identified cell classes, it is tempting to suggest that the large number of peaks that we find provide a more refined classification of cells than has been possible with other methods.  This would need to be tested by some independent measures of cell function.

\begin{figure}[t]
    \centering
    \includegraphics[width=\linewidth]{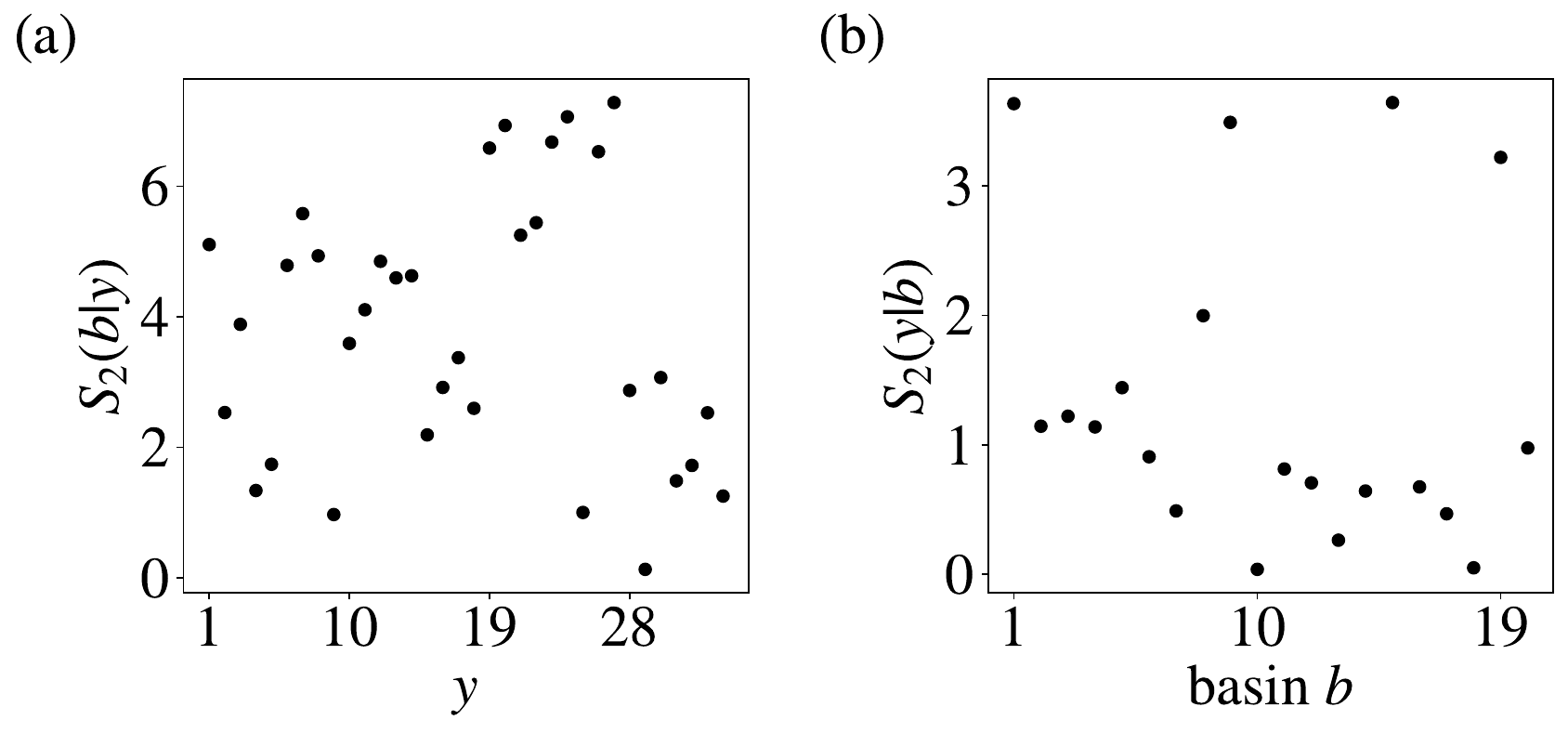}
    \caption{Correspondence basins-classes. Each class can be associated with multiple basins, while most basins are composed of cells from only one class. 
    (a). Entropy (in bits) of basin distribution for known classes, using all the 34 classes. (b).  Entropy (in bits) of the classes distribution for the 20 largest basins.
\label{fig:basins_classes_entropies}}
\end{figure}

Extending this method to include all available cells, rather than only those in the eight largest classes, we see a similar correspondence between energy basins and known cell classes. In \figref{fig:basins_classes_entropies}a
we plot the entropy of basin distribution for each known class $y$,
showing that most classes are concentrated in just a few basins. Conversely, in \figref{fig:basins_classes_entropies}b we see the entropy of basins composition for the 20 largest basins, which shows that most basins contain only a few classes.
A few basins and classes instead have a more complicated composition. Recent work has emphasized that a more rigorous treatment of noise in scRNAseq measurements implies the existence of reliable further subclasses within the clusters defined by conventional classification algorithms ~\cite{grobecker+al_24}, in agreement with the picture suggested by the statistical physics models considered here.

\section{Ising models as interpretable classifiers}

We have seen that the Ising model for gene expression patterns predicts the emergence of several local maxima in the full, high-dimensional distribution, and that these peaks correspond reasonably well to major classes or types of cells identified by other means.  This classification is completely unsupervised. 

In this section, we ignore this correspondence between energy basins and classes for a moment and further investigate the power of the maximum entropy principle. In particular, we show that we can use it to build a fully supervised cell type classifier with very high accuracy, where each class is described by a different Ising model. This suggests that the means and correlations of binarized gene expression produce models that contain all the information needed to classify cells.

\begin{figure}[b]
   \centering
   \includegraphics[width=1\linewidth]{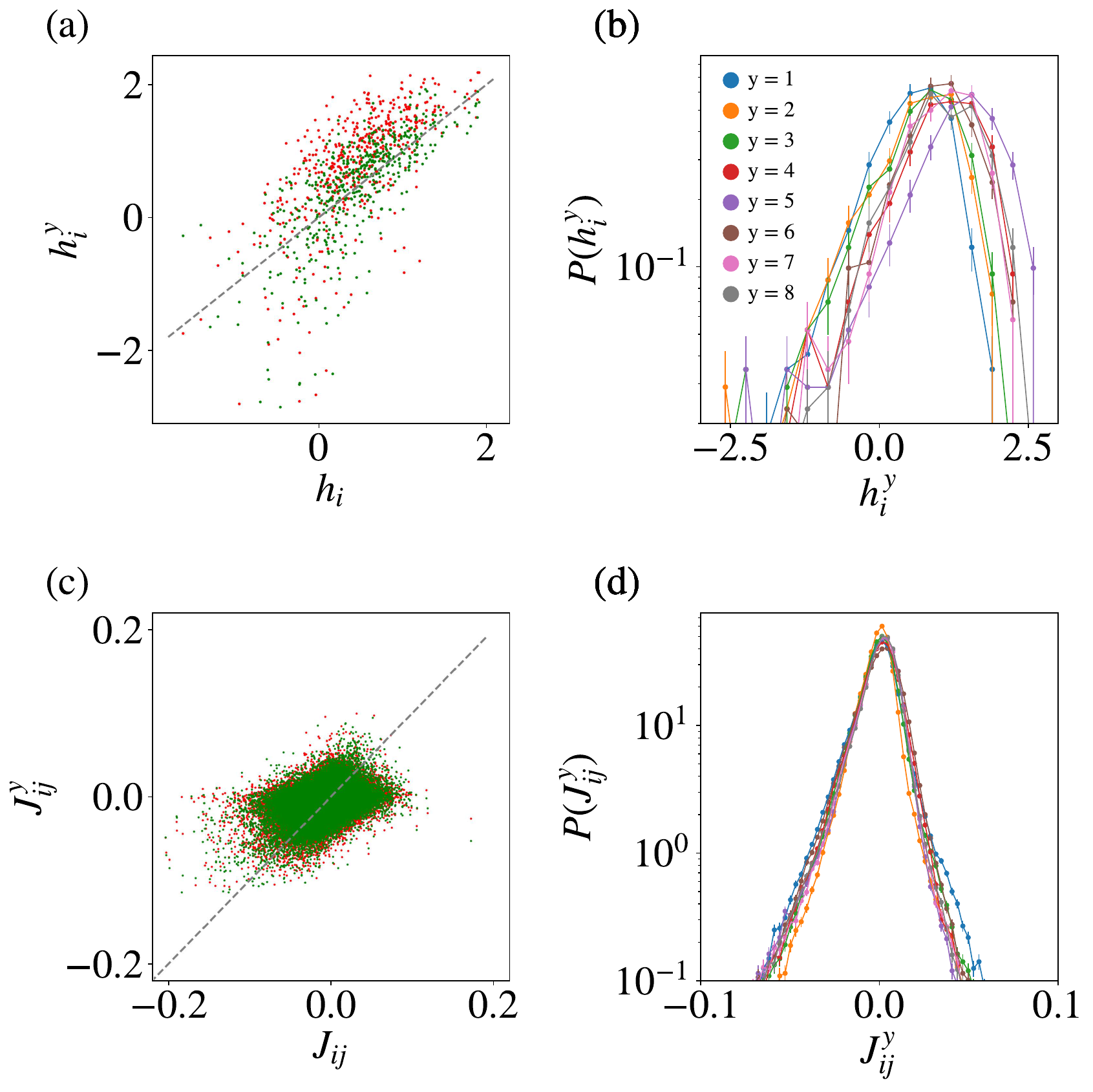}
   \caption{Comparison of the parameters of Ising models in different classes for $N=500$.  
   (a) The external fields of the overall Ising model (\eqref{me2} and \eqref{ising1}) and single model (\eqref{Pgiveny1} and \eqref{Pgiveny2}) in different classes are quite different. Classes $y=3$ and $y=4$ are shown.  
   (b) The distribution of the external fields in each of the classes $y$.
   (c) The interactions of the overall Ising model and single models defined in a single class, as in (a), are different.
   (d) The distributions of the interactions are very similar across classes.
   \label{fig:comparison_par_classes}}
\end{figure}

We can show this by looking at all cells in a particular class $y$ and evaluating the first and second moments of the binarized expression levels, $\langle \sigma_i \rangle_y$ and $\langle \sigma_i \sigma_j \rangle_y$.  Then we can follow the same construction as before to build separate maximum entropy models that are consistent with these expectation values in each class,
\begin{eqnarray}
    P(\{\sigma_i\}| y) &=& {\frac{1}{Z(y)}} \exp\left[ - E(\{\sigma_i\} ;y)\right]
    \label{Pgiveny1}\\
    E(\{\sigma_i\} ;y) &=& \sum_i h_i^y\sigma + {\frac{1}{2}} \sum_{i\ne j} J_{ij}^y \sigma_i \sigma_j .
    \label{Pgiveny2}
\end{eqnarray}
In \figref{fig:comparison_par_classes} we contrast the parameters $\{h_i^y, J_{ij}^y\}$ in these models with the $\{h_i, J_{ij}\}$ that describe the global Ising model above, with $N=500$ genes.

We see in \figref{fig:comparison_par_classes}a that each particular gene, here labelled by its value $h_i$ in the global Ising model, can have widely scattered values of $h_i^y$ in the different classes.  Interestingly the distribution of these values is similar in all classes (\figref{fig:comparison_par_classes}b).  The same pattern holds for the effective interactions $J_{ij}^y$ vs $J_{ij}$, but here the similarity of distributions across classes is even more compelling.  The fact that parameters for individual genes and pairs differ significantly across classes suggests that this could provide a basis for distinguishing among classes.

As usual mapping cells into classes should be based on the probability distribution
\begin{equation}
    P(y|\{\sigma_i\}) = {\frac{P(\{\sigma_i\}|y) P(y)}{P(\{\sigma_i\})}} .
\end{equation}
To provide a deterministic mapping of each cell $\alpha$ we can use the maximum a posteriori assignment,
\begin{equation}
    \label{eq:classifier}
    \Tilde{y}^\alpha = \underset{y}{\text{argmax }} P(y|\sigma^\alpha) .
\end{equation}
Since we have, through Eqs.~(\ref{Pgiveny1}, \ref{Pgiveny2}), an approximation to $P(\{\sigma_i\}|y) $, we can use these Ising models as the basis for classification.

\begin{figure}
    \includegraphics[width=\linewidth]{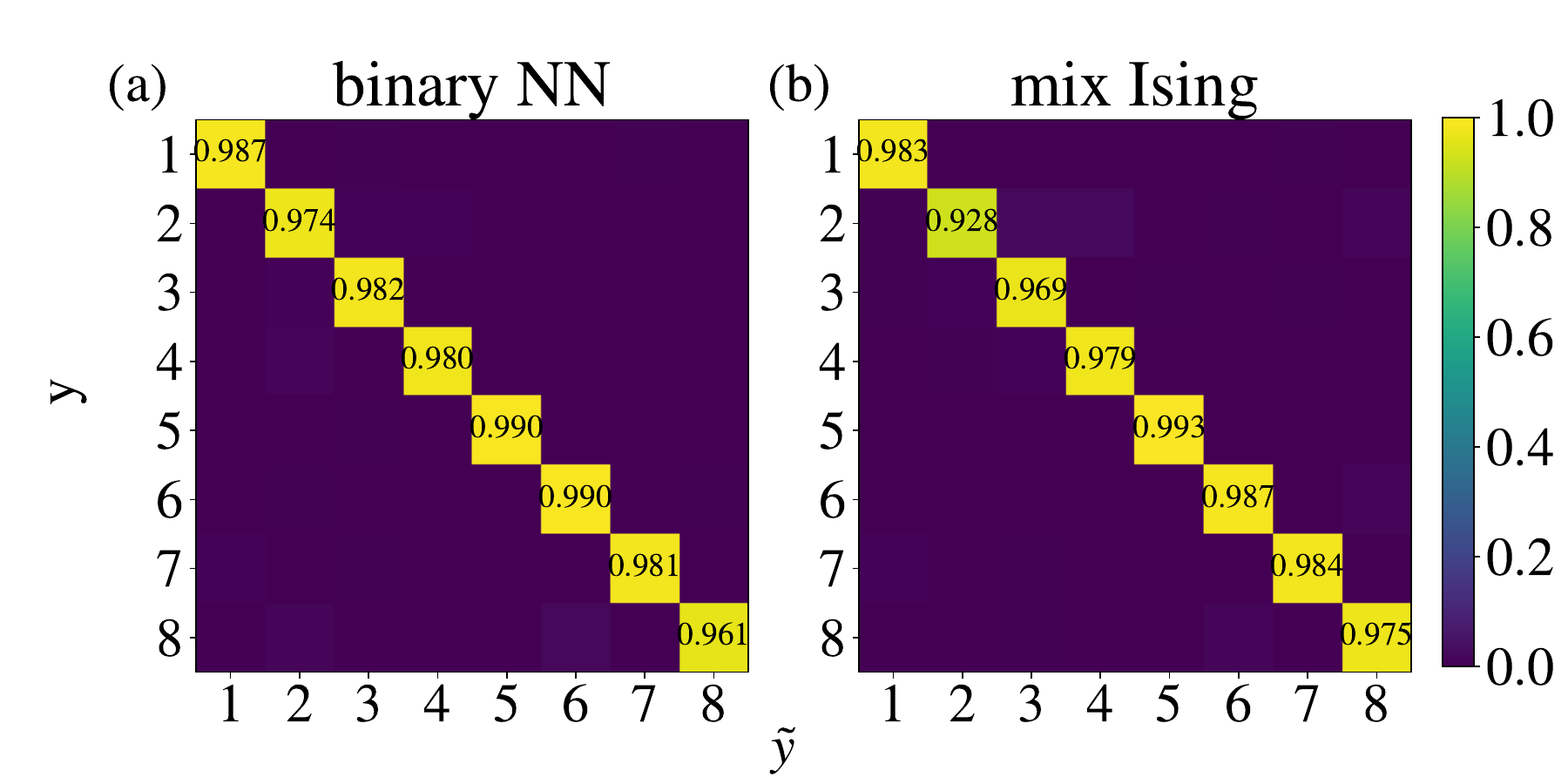}
    \caption{Confusion matrices $P(\tilde y | y)$ for classification based on $N=500$ genes.  (a) Performance of the neural network from \figref{fig:scheme_NN}.  The weighted mean of the diagonal gives the overall performance shown in \figref{NNperformance}.   (b) Classification based on the Ising model approximation to $P(\{\sigma_i\}|y )$.
        \label{fig:accuracy}}
\end{figure}

In \figref{fig:accuracy} we show the confusion matrix generated by this ``Ising classifier'' compared with that of the neural network (\figref{fig:scheme_NN}).

Strikingly, the performance is essentially the same,  with the mixture of Ising models in some cases doing better than the neural network.  
This suggests that the existing classification depends essentially on the presence or absence of mRNA species and their pairwise correlations, and hence that Ising models constitute an interpretable, physically motivated basis for supervised classification.

\section{Discussion}
The development of advanced sequencing techniques, such as scRNA-seq and later MERFISH, has revolutionized cell classification. These techniques have provided more objective and well founded criteria for classifying large groups of cells based on gene expression data, in addition to the previous purely morphological and functional classification.
However, most existing methods still require ad hoc parameter tuning, such as setting the number of clusters or the distance threshold, and involve preprocessing procedures and arbitrary choices to align with known biological information.

Importantly, our work provides unsupervised classification without the need for external parameters or a predefined number of clusters. Different groups of cells emerge naturally from the energy basins of the reconstructed probability distribution. These groups are in good agreement with previously known cell types, and in some cases even show a finer structure, as shown in Figure 16. 
Furthermore, this work adds to the growing number of examples showing that the maximum entropy principle is a powerful tool for dealing with large data sets in a wide range of living systems. 
In particular, we define binary variables representing the presence or absence of mRNA species and show that this binary information is sufficient to distinguish cell types.  The proposed maximum entropy model consistent with the mean and correlations of these binary variables---an Ising model---successfully reproduces higher order statistics such as the third and fourth connected moments and the distribution of summed expression levels without explicitly enforcing them.

In the future, this classification method could be applied beyond neurons, providing new insights into cell types and their development.

\acknowledgments
Work at Princeton was supported in part by the  National Science Foundation, through the Center for the Physics of Biological Function (PHY--1734030), and by fellowships from the John Simon Guggenheim Memorial Foundation (WB), the Schmidt Science Fellowship (TG), and the Simons Foundation (WB). Work at the Flatiron Institute is supported by the Simons Foundation (LS). The work of CGC was performed in part at Aspen Center for Physics, which is supported by
National Science Foundation (PHY-2210452).

\appendix

\section{Ising models for $N=100$ and $N=200$}
\label{app:extra plots}

\begin{figure}[b]
    \centering
    \includegraphics[width=1\linewidth]{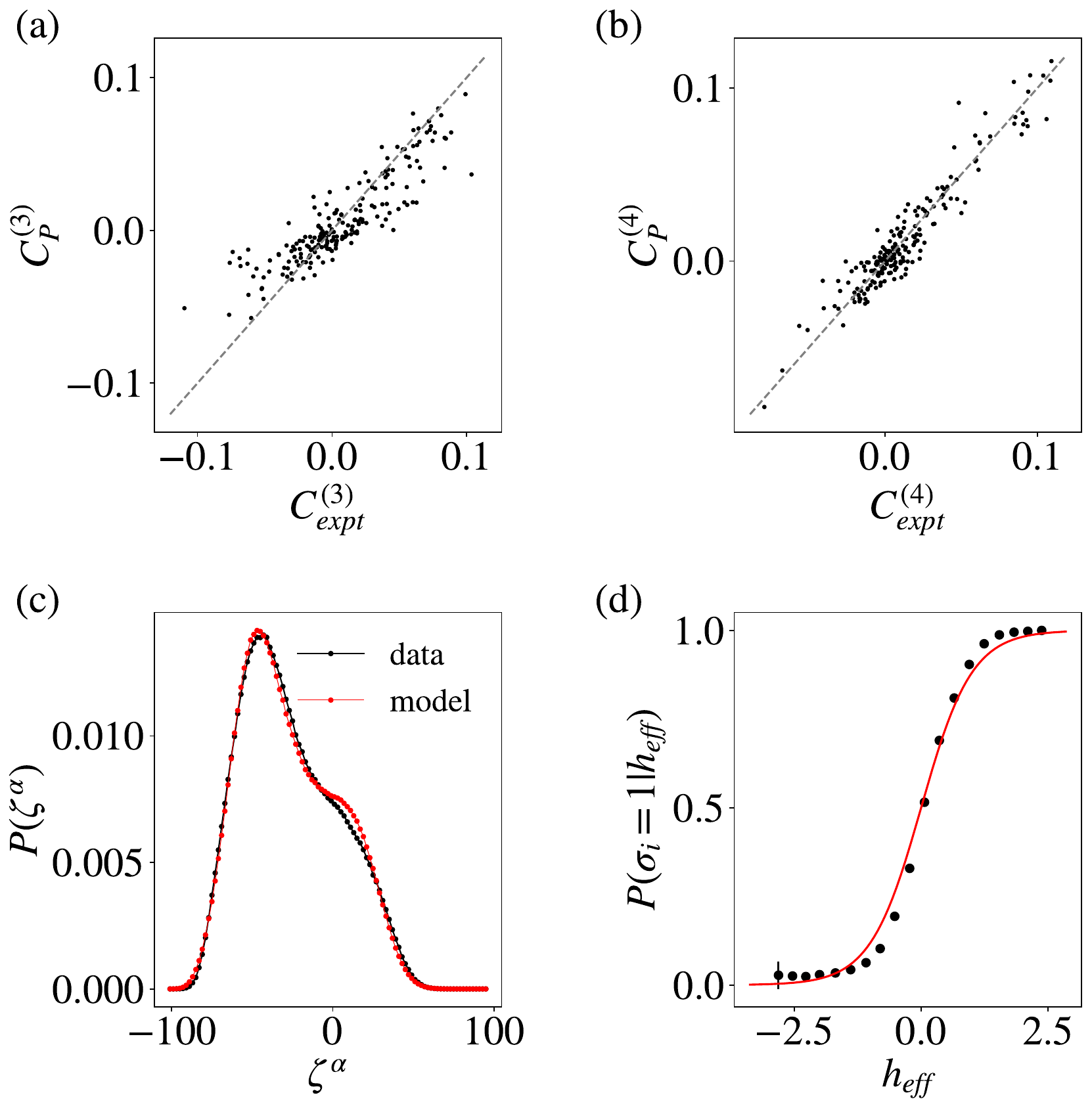}
    \caption{Testing the Ising model for a selection of $N=100$ genes, following \figref{fig:ising_higher}.
        (a) Third-order connected moments. Comparison between data moments and Ising model predictions for 500 random combinations. The diagonal represents perfect agreement between data and model predictions.
        (b) Fourth-order connected moments for 500 random combinations.
        (c) Sum distribution (\eqref{z_binary_def}): comparison between data (black) and Ising model (red).
        (d) Probability of one variable given the state of all the other genes.  The red curve is given by \eqref{eq:heff}.}
    \label{test100}
\end{figure}

\begin{figure}
    \centering
    \includegraphics[width=1\linewidth]{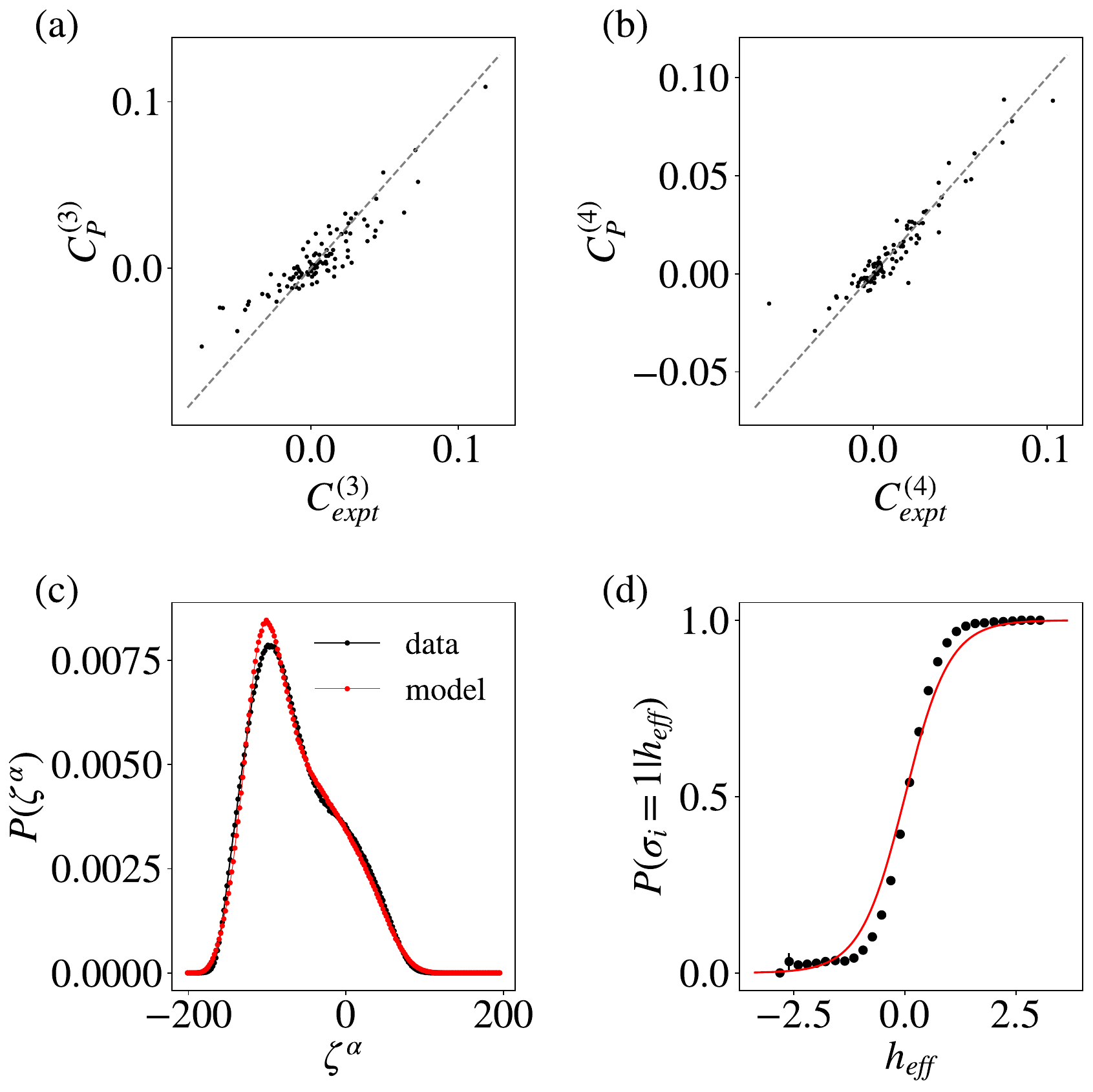}
    \caption{Testing the Ising model for a selection of $N=200$ genes, following \figref{fig:ising_higher}.
        (a) Third-order connected moments. Comparison between data moments and Ising model predictions for 500 random combinations. The diagonal represents perfect agreement between data and model predictions.
        (b) Fourth-order connected moments for 500 random combinations.
        (c) Sum distribution (\eqref{z_binary_def}): comparison between data (black) and Ising model (red).
        (d) Probability of one variable given the state of all the other genes.  The red curve is given by \eqref{eq:heff}.}
    \label{test200}
\end{figure}

In the main text we assess the performance of a maximum entropy model built to match the mean and covariance matrices of $\{\sigma_i\}$ for all $N=500$ genes (\figref{fig:ising_higher}).  Here we make the same comparisons between theory and experiment for models based on a selection of $N=100$ (\figref{test100}) and $N=200$ (\figref{test200}) genes.

Agreement between theory and experiment is generally good across the range of $N$ considered here.  There are some hints that agreement is better at smaller $N$, but this is a small effect.  Better agreement would require matching more expectation values, and it is not clear which ones are the most informative.

\bibliography{main.bib}

\end{document}